\documentclass[conference]{IEEEtran}

\PassOptionsToPackage{dvipsnames*,svgnames,x11names}{xcolor}

\usepackage{cite}
\usepackage{graphicx}
\usepackage{tikz}
\usepackage{soul}
\usepackage{amsmath}
\usepackage{amssymb}
\usepackage{comment}
\usepackage[normalem]{ulem}
\usepackage[htt]{hyphenat}
\usepackage{braket}
\usepackage[ruled,vlined]{algorithm2e}
\usepackage{listings}
\usepackage{subcaption}
\usepackage{tabularx}
\usepackage{booktabs}
\usepackage{multirow}
\usepackage{makecell}
\usepackage{hyperref}
\usepackage{algpseudocode}

\usepackage{tcolorbox}
\usepackage{mdframed}
\usepackage{colortbl}
\usepackage{grumble}
\usepackage{arydshln}

\usepackage[inline]{enumitem}
\setlist{noitemsep,topsep=0pt,parsep=0pt,partopsep=0pt}

\usepackage[subtle]{savetrees}

\usepackage{setspace}
\usepackage[margin=5pt,font={stretch=0.9}]{caption}

\newcommand{\projecttitle}{MCMit}

\title{\projecttitle: Hardware-Software Co-Design for Mid-Circuit Measurement Error Mitigation}

\author{\parbox{0.95\textwidth}{\centering
{\sublargesize Emmanouil Giortamis, Felix Gust, Aleksandra Świerkowska, Sandra Stankovic, Innocenzo Fulginiti, Yanbin Chen, Xiaorang Guo, Benjamin Lienhard\textsuperscript{$\dagger$}, Martin Schulz, and Pramod Bhatotia}\\[6pt]
\normalsize Technical University of Munich, Munich, Germany\\
\textsuperscript{$\dagger$}Also with the Walther-Meißner-Institut and the Munich Center for Quantum Science and Technology, Munich, Germany\\
Email: firstname.lastname@tum.de}}

\definecolor{mycolor}{HTML}{0000FF}

\begin{document}



{\microtypesetup{tracking=false}\maketitle}

\begin{abstract}
Distributed Quantum Computing (DQC) and Quantum Error Correction (QEC) rely on dynamic circuits that include Mid-Circuit Measurements (MCMs) and classical feedback. These operations present a major bottleneck: MCMs suffer from high error rates that lead to real-time branching errors, while MCM and classical feedback latencies amplify decoherence errors. Current hardware controllers, qubit-state discriminators, and software error mitigation techniques fail to address these challenges holistically.

We propose \projecttitle{}, a hardware-software co-design to mitigate branching and latency-induced errors. \projecttitle{} introduces a scalable, constant-latency multi-control branch instruction for faster classical feedback and two qubit-state discriminators, a transformer, and a CNN, with high accuracy even under short measurement durations. On the software side, static MCM elimination and stochastic branching complement the hardware by mitigating residual branching errors that persist despite hardware improvements.

We implement \projecttitle{} on Qubic and evaluate it using experimentally extracted QPU readout traces. Our branch instruction reduces feedback latency by up to 70\%, improving circuit depths by up to $7\times$ over Qubic. Our CNN discriminator achieves up to 62\% higher accuracy for short measurement durations than the baselines, driving $1.2-9.4\times$ lower logical error rates in QEC. Last, our software mitigation improves fidelity by 18--30\% over baseline methods in DQC workloads.

\end{abstract}

\pagestyle{plain}

\section{Introduction}
\label{introduction}

\myparagraph{Context and motivation}
Current Noisy Intermediate-Scale Quantum (NISQ) processors face two primary obstacles hindering progress towards quantum advantage: inherent operational noise and limitations in controlling large qubit counts (scale)~\cite{preskill2018nisq}. Overcoming noise fundamentally requires the implementation of Quantum Error Correction (QEC) \cite{knill1997theory, gidney2021how, preskill1997faulttolerantquantumcomputation}, while transcending scale limitations necessitates Distributed Quantum Computing (DQC) architectures that interconnect multiple processors \cite{caleffi2024distributed, bravyi2022the}. Crucially, both QEC and DQC depend on dynamic quantum circuits, which incorporate Mid-Circuit Measurements (MCMs) and subsequent classical feedback \cite{corcoles2021exploiting, graham2023Midcircuit, wan2019quantum}. These operations are indispensable for core operations, such as syndrome measurements in QEC and the classical communication inherent in quantum teleportation protocols, essential for DQC \cite{wu2022autocomm, wuqucomm2023}.

Despite their utility, MCMs, particularly in superconducting architectures \cite{ibm-devices}, exhibit both high error rates and substantial latency, often representing the slowest and most error-prone operations in a quantum circuit \cite{tannu2019mitigating, das2021jigsaw, patel2020experimental}. Furthermore, the subsequent classical feedback required for dynamic circuits introduces additional, often unaddressed, latency that ranks second only to the measurements themselves, further hindering performance, as we show in Table \ref{tab:error_rates}.

The above limitations inherent in MCMs present significant challenges for DQC and QEC. Firstly, MCM errors (bitflips) manifest as \textit{real-time branching errors}. 
Branching errors occur when an MCM error causes the classical control logic to deviate from the intended execution path, triggering an incorrect conditional quantum operation. Unlike terminal measurement errors, which can be mitigated post-execution via statistical corrections on the final outcome distribution \cite{das2021jigsaw, dangwal2024varsaw, tannu2019mitigating}, branching errors propagate irrecoverably: an MCM error in a teleportation protocol applies the wrong Pauli correction, while a syndrome MCM error in a QEC cycle can trigger a recovery gate that worsens rather than corrects the logical error.

Secondly, the combined latency of MCMs and classical feedback is detrimental; during teleportation, this delay allows the receiver qubit to decohere before the necessary correction can be applied, reducing fidelity \cite{baumer2024efficient, carrera2024combining}. In QEC cycles, this latency extends the time during which logical errors can accumulate between syndrome measurement rounds, directly competing against the code's ability to correct errors. We quantitatively motivate these claims in \S~\ref{motivation}. 

\begin{figure*} [t]
    \centering
    \includegraphics[width=0.9\textwidth]{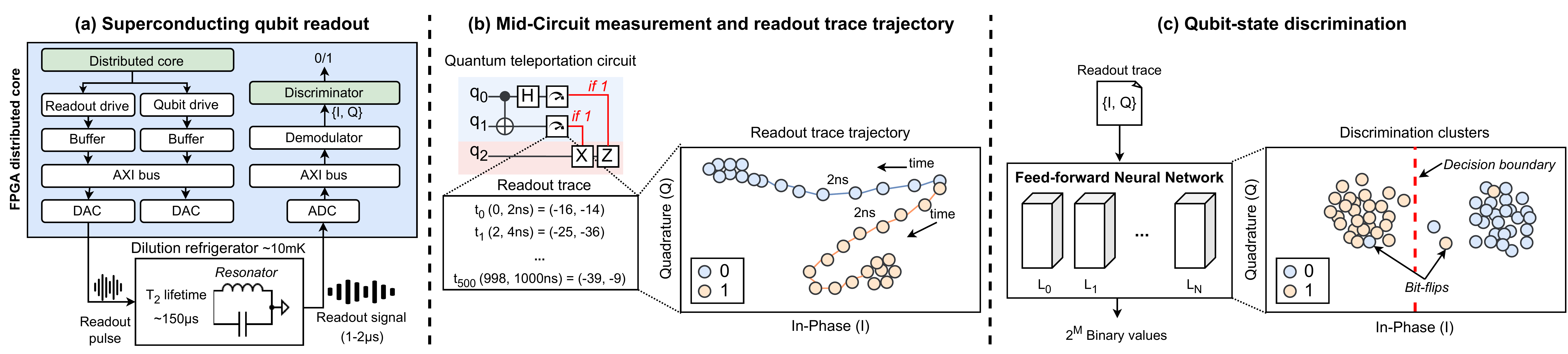}
    \caption{Superconducting readout (\S~\ref{background:readout_pipeline}).
    \textit{
    \textbf{(a)} Superconducting qubit readout pipeline on an FPGA controller.
    \textbf{(b)} The MCM in a teleportation circuit can produce a readout trace of 1$\mu s$: 500 samples spaced 2ns apart.
    \textbf{(c)} The readout trace is input to a Feed-forward Neural Network comprising N layers. The network discriminates 0 and 1 using a decision boundary that separates the states.
    }}
    \label{fig:background}
\end{figure*}

\myparagraph{Research gap}
Unfortunately, current methods for enhancing the fidelity and speed of MCMs and feedback operations are not only \textit{individually limited}, but are developed in isolation, with \textit{no cross-layer co-design} that accounts for their mutual dependencies.
First, real-time FPGA controllers \cite{xu2021qubic} lack native support for complex multi-qubit conditional logic, essential for protocols in DQC and QEC \cite{baumer2024efficient, xu2026distilling}, leading to feedback \textit{latency that scales with circuit complexity} rather than remaining constant.
Second, qubit-state discriminators \cite{maurya2023scaling, vora2024ml} are optimized for \textit{long readout} durations, making them ill-suited for the \textit{short durations} required by fast QEC cycles, both to avoid decoder backlog under the 1$\mu$s latency budget \cite{terhal2015quantum} and to sustain high repetition rates, where their accuracy degrades significantly, directly increasing the logical error rate. Third, existing software error mitigation operates purely post-execution \cite{das2021jigsaw, dangwal2024varsaw, tannu2019mitigating}, adjusting final measurement distributions after the fact, being fundamentally oblivious to real-time branching errors that have already propagated through the circuit, and is designed with no knowledge of hardware calibration data or discriminator confidence. Crucially, none of these systems is aware of the others.

\statementLeftAccent{Research Question}{orange}{How can FPGA controllers, qubit-state discriminators, and software error-mitigation techniques be co-designed such that each layer exploits and compensates for the others, collectively removing the latency and error bottleneck of MCMs in DQC and QEC?}

\myparagraph{Our approach}
To answer this question, we propose \projecttitle{}, the first hardware-software co-design to mitigate errors and latency associated with MCMs and classical feedback. On the hardware side, \projecttitle{} \textit{enables faster classical feedback} by introducing a scalable multi-control branch instruction that has constant-latency w.r.t. the number of qubits as input. MCMit features two powerful qubit-state discriminators, a transformer, and a CNN that achieve high accuracy even with short readout durations, thereby \textit{enabling faster measurements}. Complementing the hardware, \projecttitle{}'s software \textit{mitigates MCM-induced errors} by employing static analysis to eliminate MCMs and their subsequent feedback where feasible, thereby reducing associated errors and delays, and utilizing stochastic branching to mitigate the impact of branching errors caused by MCM errors.

\projecttitle{}'s design directly addresses key technical hurdles in MCM error mitigation across the stack, with each layer explicitly informed by the others. Specifically, the utility of fast branching depends on the discriminator's ability to produce reliable outcomes quickly. Short readout durations, necessary to keep the feedback window narrow, demand models that capture temporal signal features with limited data, motivating our transformer discriminator. In turn, even a fast and accurate hardware stack cannot eliminate MCM errors entirely; the software layer closes this gap by consuming hardware-provided confusion matrices to inform stochastic branching decisions at compile time, and by applying static MCM removal to reduce the number of branch instructions the controller must execute at runtime, directly feeding back into hardware latency. Together, these layers form a closed loop: the discriminator enables fast feedback, the branch instruction exploits it, and the software mitigates what hardware leaves unresolved.

We implement the \projecttitle{} controller on top of the open-source Qubic 2.0 framework \cite{xu2021qubic} and deploy it on the Xilinx Alveo XCU280 FPGA card, the \projecttitle{} transformer and CNN using PyTorch and Torchvision \cite{paszke2019pytorch}, and train them with an Nvidia A40 GPU, and last, the \projecttitle{} software in Python using the Qiskit framework \cite{javadiabhari2024qiskit}.

We evaluate \projecttitle{} using \textit{real} readout traces from a five-qubit QPU described by Lienhard et al. \cite{lienhard2022deep}, four micro-benchmarks that utilize MCMs and classical feedback, and two high-level applications, MECH \cite{hezi2024MECH} for DQC and the surface code for QEC \cite{fowloer2012surface}. On hardware, the \projecttitle{} branch instruction incurs up to 70\% lower feedback latency, leading to 37\%-57\% higher fidelity and 7$\times$ depth improvement, compared to Qubic. The \projecttitle{} CNN outperforms Qubic's ML discriminator, QubiCML, by 11.6\% and achieves up to 62\% higher accuracy and 30.9\% lower circuit depth for short readout durations (250ns-500ns) compared to the state-of-the-art qubit state discriminator HERQULES \cite{maurya2023scaling}.  This leads to 1.2$\times$-9.4$\times$ lower logical error rate for the surface QEC code under current circuit-level noise.
Last, on the software side, \projecttitle{}  achieves 18-30\% higher fidelity compared to Qiskit mthree \cite{nation2021scalable} and unmitigated circuits, respectively.

\projecttitle{} is available at \url{https://doi.org/10.5281/zenodo.21358489}.

\myparagraph{Contributions} We make the following contributions:

\begin{itemize}
    \item We introduce the first holistic hardware-software co-design for MCM and classical feedback error mitigation. 
    \item We present an FPGA-based controller that implements a real-time and scalable branch instruction for constant-latency multi-qubit classical feedback (\S~\ref{controller}). 
    \item We present two qubit-state discrimination models designed for enabling high-accuracy short-duration readout (\S~\ref{state-discrimination}).
    \item We present software that simplifies dynamic circuits, mitigating their associating errors and latency, and suppresses branching errors stochastically (\S~\ref{compiler}).
\end{itemize}

\begin{figure*} [t]
    \centering
    \includegraphics[width=\linewidth]{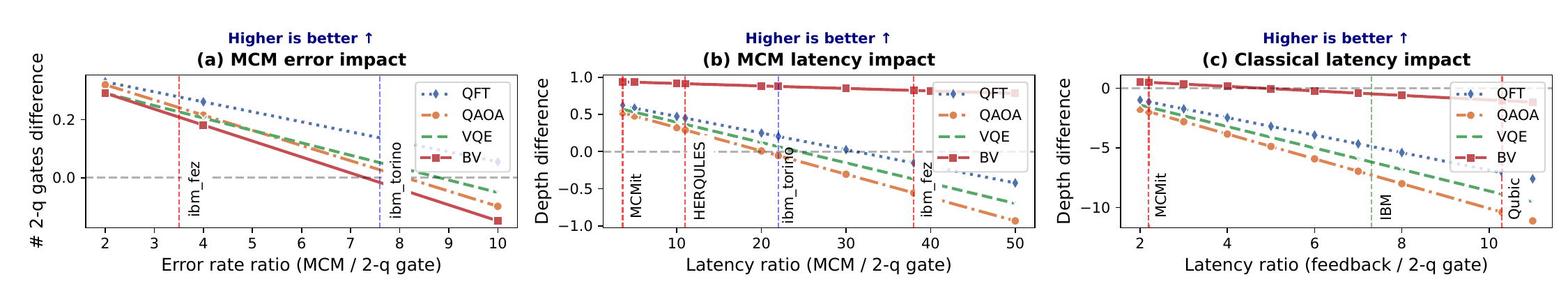}
    \caption{MECH \cite{hezi2024MECH} performance analysis. \textit{
    \textbf{(a)} Impact of MCM error rate (normalized to 2-q gate error rate) on MECH's effective 2-q gate count.
    \textbf{(c)} Impact of classical feedback latency as a ratio to the 2-qubit gate latency. There is a linear performance improvement with MCM error and (classical) latency reduction.
    }}
    \label{fig:mech_analysis}
\end{figure*}

\section{Background}
\label{background}

In this section, we detail the superconducting qubit readout pipeline, the core process executed during each MCM, and then contextualize the importance of MCMs as pivotal operations enabling both scalability through DQC and fault-tolerance via QEC.

\subsection{Qubit Readout in Quantum Architectures}
\label{background:readout_pipeline}

To set the stage, we must first understand the qubit readout pipeline.

\myparagraph{Quantum architecture readout pipeline [Fig.~\ref{fig:background} (a)]}
Qubit-state readout in superconducting QPUs is handled by FPGA control hardware \cite{xu2021qubic, vora2024ml, yang2022FPGA, messaoudi2020a}. Fig. \ref{fig:background} (a) shows the pipeline of the Qubic open-source QPU control framework \cite{xu2021qubic}, where the core hardware components are shown in green. We refer interested readers to the remaining components in \cite{heinsoo2018rapid, lienhard2022deep, acharya2023multiplexed, heinsoo2018rapid, diguglielmo2025end, delaney2022superconducting}.
The distributed core instructs the readout drive to perform a measurement, and the readout pulse information is sent over the AXI bus to the FPGA's DAC. The pulse is converted into an analog microwave-frequency pulse and transmitted to the qubit's resonator within the dilution refrigerator. The resonator undergoes a qubit-state-dependent phase shift. The resulting transmitted signal is first demodulated and then digitized. The result is a \textit{readout trace} that is the input to a qubit-state discriminator, which generally classifies the trace as 0 or 1.

\myparagraph{Readout traces and trajectories [Fig. \ref{fig:background} (b)]}
A readout trace is a sequence of In-Phase (I) and Quadrature (Q) $\{I,Q\}$ points forming a 2D plane, and shows the time evolution (trajectory) of the readout signal and the qubit's final state \cite{magesan2015machine, lienhard2022deep}. Fig. \ref{fig:background} (b) shows a readout trace with length 1000ns: 500 samples spaced 2ns apart. On the right, we show an example readout trajectory, where the x-axis represents the I component and the y-axis represents the Q component of the trace points.

\myparagraph{Qubit-state discrimination [Fig. \ref{fig:background} (c)]}
Qubit-state discrimination outputs the final binary measurement outcome. To perform this task, state-of-the-art discriminators leverage Feed-forward Neural Networks (FNNs) \cite{vora2024ml, lienhard2022deep, maurya2023scaling}, which achieve high accuracy and can be implemented efficiently on FPGA controllers. Fig. \ref{fig:background} (c) shows an example FNN with $N$ layers (input, output, and $N-2$ hidden layers) that produces a $2^M$-bit output, with $M$ denoting the number of qubits. The output dimension depends on the design, where some designs output multi-qubit states \cite{maurya2023scaling} while others output single-qubit states \cite{vora2024ml}. Last, these networks create a decision boundary for discriminating states (red dashed line in Fig. \ref{fig:background} (c)) \cite{diguglielmo2025end, vora2024ml}.

\subsection{MCMs for Scalability and Fault Tolerance}
\label{background:DQC}

\myparagraph{Scalability: Distributed Quantum Computing (DQC)}
Monolithic QPUs face key scalability limitations that prohibit the fabrication of QEC-scale chips \cite{staszewski2021cryo, krinner2019engineering, valenzuela2006microwave}. This leads to the use of distributed architectures, such as chiplet designs, which connect multiple QPUs via short-range couplers \cite{ibm-remote-gate-cost}, and DQC setups utilizing quantum networks \cite{bravyi2022the, ang2024arquin}, that utilize components such as quantum repeaters, quantum switches, and quantum transduction \cite{barral2025review, kozlowski2019towards}.

\myparagraph{DQC communication primitives}
Communication between two QPUs in DQC setups is based on two main communication primitives, which utilize MCMs and classical feedback operations. First, \textit{quantum teleportation} enables communication between distant QPUs using one remote 2-qubit operation, and at least \textit{two} MCMs and \textit{two} classical feedback operations, with variants such as gate and state teleportation \cite{zeilinger2000quantum, pirandola2015advances, wu2022autocomm}.  Second, \textit{constant-depth long-range entanglement} consumption involves preparing and using entangled states (e.g., GHZ) at constant circuit depth using dynamic circuits, to apply remote operations efficiently and reduce inter-QPU communication \cite{hezi2024MECH, baumer2024efficient, kang2025teleporting, waring2026robust}. Crucially, preparing these states in constant depth requires non-trivial classical control flow: MCM outcomes across multiple qubits must be jointly processed via controller-level XOR or majority voting operations to propagate correction logic, before the corresponding feedback gates can be applied \cite{baumer2024efficient, waring2026robust}.

\myparagraph{Fault tolerance: Quantum Error Correction (QEC)}
Achieving quantum advantage is widely believed to require fault-tolerant quantum computers, making QEC indispensable for overcoming the intrinsic fragility of qubits and operational noise \cite{fowloer2012surface, gidney2021how, preskill1997faulttolerantquantumcomputation}. QEC works by redundantly encoding a logical qubit across multiple physical qubits, such that common physical errors transform the state into distinguishable error subspaces \cite{Nielsen_Chuang_2010, Roffe_2019}. Errors are identified by performing MCMs on ancilla qubits, extracting syndrome information that reveals which error subspace the system occupies without collapsing the logical state \cite{decoding_2023}; these MCMs must be performed cyclically to detect and prevent error propagation \cite{Roffe_2019}.
As such, the fidelity and speed of each MCM cycle critically dictate both the quality of the extracted syndrome information and the time available for further errors to accumulate \cite{harper2025characterisingfailuremechanismserrorcorrected}.

\section{MCM Errors in DQC and QEC Applications}
\label{motivation}

To quantitatively motivate \projecttitle{}, we characterize MCM and feedback errors and latencies on state-of-the-art superconducting processors \cite{ibm-devices}, and evaluate their impact on DQC scalability via MECH \cite{hezi2024MECH}, a compiler for chiplet architectures, and on fault tolerance via the surface code under realistic IBM processor assumptions.

\begin{table}[t]
\footnotesize
\caption{IBM Heron and Nighthawk processors' properties: Absolute values and normalized to those of a 2-qubit gate.\protect\footnote{Absolute values and normalized ratios vary by device and by calibration day; the reported ranges are representative rather than exhaustive bounds.}}

\begin{center}
\begin{tabular}{ |c|c|c| }
 \hline
 \bf{Operation} & {\bf Error rate} & {\bf Duration} \\ \hline \hline
 {1-qubit gate} & 2e-4 ($\leq 0.1 \times$) & 32ns ($0.35-0.45\times$) \\ \hline
 {2-qubit gate} & 1.6e-2-2.7e-3 ($1\times$) & 68-88ns ($1\times$)  \\ \hline
 {Measurement} & 4e-3-3e-2 ($\mathbf{2-12\times}$) & 1.5-2.5$\mu s$ ($\mathbf{23-38\times}$) \\ \hline
  {Remote 2-q gate \cite{ibm-remote-gate-cost}} & 3.5e-2 ($10-18\times$) & 235ns ($2.6-3.4\times$) \\ \hline
Classical feedback \cite{xu2021qubic}  \cite{carrera2024combining} & 0 & {205-701ns} ($\mathbf{3-10\times}$) \\
  \hline

\end{tabular}
\end{center}
\label{tab:error_rates}
\end{table}

\subsection{Readout Errors and Latency}
\label{motivation:readout_errors}

In the following, we compare the readout error rates, as well as the readout and classical feedback latencies, to the other operations supported by state-of-the-art IBM QPUs.

\myparagraph{Readout errors}
In currently available superconducting QPUs, qubit readout is the most error-prone operation \cite{patel2020experimental, tannu2019mitigating, ibm-devices}, as shown in Table \ref{tab:error_rates}.\footnote{Exact ratios depend on the specific device and calibration day; the reported ranges are representative rather than exhaustive bounds.} On current IBM Heron and Nighthawk processors, readout error rates are typically $2-12\times$ higher than those of two-qubit gates, depending on the specific QPU \cite{ibm-devices}. Specifically, readout errors arise from \textit{relaxation} during the relatively long measurement window \cite{picot2008role, thorbeck2024readout, mallet2009single}, \textit{crosstalk} in frequency-multiplexed setups \cite{ryan2015tomography, chen2012multiplexed, lienhard2022deep, walter2017rapid}, qubit \textit{excitations} induced by the readout pulse \cite{walter2017rapid, chen2016measuring}, and imperfect suppression of \textit{environmental noise} despite signal amplification and filtering \cite{maclin2015a, sete2015quantum}.

\myparagraph{Readout latency}
Currently, readout is optimized for maximum \textit{terminal} measurement fidelity, not MCM fidelity. As such, measurements are also the slowest operation, taking approximately $20-40\times$ longer than a two-qubit gate on IBM Heron processors \cite{ibm-devices}. Concretely, readout length spans 1560-2584ns, while a two-qubit gate only spans 68-88ns. Interestingly, this latency ``wall'' renders dynamic circuits non-optimal compared to unitary circuits or post-selection, despite the use of dynamic decoupling \cite{baumer2024efficient}.

\myparagraph{Classical feedback latency}
Interestingly, classical feedback operations are the \textit{second slowest operation} \cite{wuwei2025artery}. In the seminal Google quantum error correction experiment, feedback takes 160ns, in contrast to 32ns two-qubit gates and 500ns readout (5$\times$ and 32\%, respectively) \cite{google2023suppressing}, and IBM reports a 500ns feedback latency in their DQC experiment {(20-30\% of measurement duration)} \cite{carrera2024combining}. Last, the state-of-the-art open-source quantum controller framework, Qubic \cite{xu2021qubic}, supports classical feedback for 205-701ns latency ($3\times-10\times$), depending on the complexity of the control flow. Notably, classical feedback becomes first-order in \projecttitle{}'s target regimes, where feedback latency scales with the number of jointly-processed MCMs and accumulates across rounds and patches.

\takeaway{
MCM errors and the combined latency of measurement and feedback severely hinder the performance of dynamic circuits, creating a major bottleneck for QEC and DQC applications.
}

\subsection{MCM Impact on DQC}
\label{motivation:dqc}

We begin our quantitative motivation with MECH \cite{hezi2024MECH}, the state-of-the-art compiler for chiplet architectures.

\myparagraph{Experimental setup}
Without available DQC setups, the MECH compiler was originally evaluated at the compilation level using two ``effective'' metrics: effective CNOT count and effective depth. In this framework, MCMs and remote CNOTs are normalized to the cost of a local CNOT gate, scaling by error rate for gate count and by latency for depth. This normalization means that as MCM cost increases, the reported metric grows proportionally even when physical gate counts remain constant. We extend this evaluation with empirical error rates and latencies from IBM Heron devices \cite{ibm-devices} for a more realistic assessment of execution-time trade-offs in near-term chiplet-based architectures.

\myparagraph{Problem \#1: 2-q gate increase from MCM errors}
Fig. \ref{fig:mech_analysis} (a) shows the impact of MCM errors evaluated as the increasing ratio of the MCM to 2-q gate error rates. The y-axis shows MECH's performance difference as effective 2-q gate count difference, where $y>0$ means performance improvement. Here, a remote 2-q gate is $10\times$ noisier than a local 2-q gate. We also plot the MCM/2-q error rate ratio of the \texttt{ibm\_fez} and \texttt{ibm\_torino} QPUs \cite{ibm-devices}. Notably, lower MCM error rates linearly improve MECH performance, since there are $6.4\times$ more MCMs than remote 2-q gates ($5.1-8.6\times$), effectively contributing $ 1.2- 2.3\times$ more errors compared to remote 2-q gates. 

\myparagraph{Problem \#2: Circuit depth increase from MCM latency}
Fig. \ref{fig:mech_analysis} (b) shows the impact of the increasing MCM/2-q gate latency ratio. The y-axis shows MECH's post-compilation depth difference, where every operation is normalized to a 2-q gate, and $y>0$ means lower depth. We assume an MCM/2-q gate error ratio of 2, which is better than ibm\_fez in (a). Again, MCM latency reduction linearly increases MECH's depth performance. Interestingly, by enabling fast readout ($<$750ns), the latency impact of MCMs is lower; for instance, \projecttitle{} with high-accuracy 250ns readout would improve MECH's depth by 2.4$\times$ and 11$\times$ compared to \texttt{ibm\_torino} and \texttt{ibm\_fez}, respectively.

\myparagraph{Problem \#3: Circuit depth increase from classical latency}
Last, Fig. \ref{fig:mech_analysis} (c) shows the impact of increasing classical feedback/2-q gate latency ratio. We assume a MCM/2-q gate latency ratio of 10 (lower than ibm\_torino in (b)). To our surprise, classical latency was considered to be \textit{zero} in the original MECH evaluation and thus was \textit{not} accounted for. We plot the feedback latencies of Qubic \cite{xu2021qubic}, and those of IBM \cite{baumer2024efficient, carrera2024combining} and \projecttitle{}. As latency increases (decreases), MECH's performance deteriorates (improves) in a linear fashion. Notably, by enabling constant-latency feedback, \projecttitle{} can achieve a depth $\sim$7$\times$ lower than Qubic.

\begin{takeaway} {
MCM errors and latencies detrimentally impact the performance of DQC compilers, since the combined measurement and feedback latency increases the effective circuit depth. However, \projecttitle{} can lower the depth by 11$\times$ and 7$\times$ by enabling faster readout and implementing constant-latency feedback, respectively.
}

\end{takeaway}

\subsection{MCM Impact on QEC}
\label{motivation:qec}

{We   evaluate the impact of readout error rate and duration on the surface QEC code used in the Google Sycamore experiment \cite{google2023suppressing}. }

\begin{figure}[t]
   \centering
   \includegraphics[width=\linewidth]{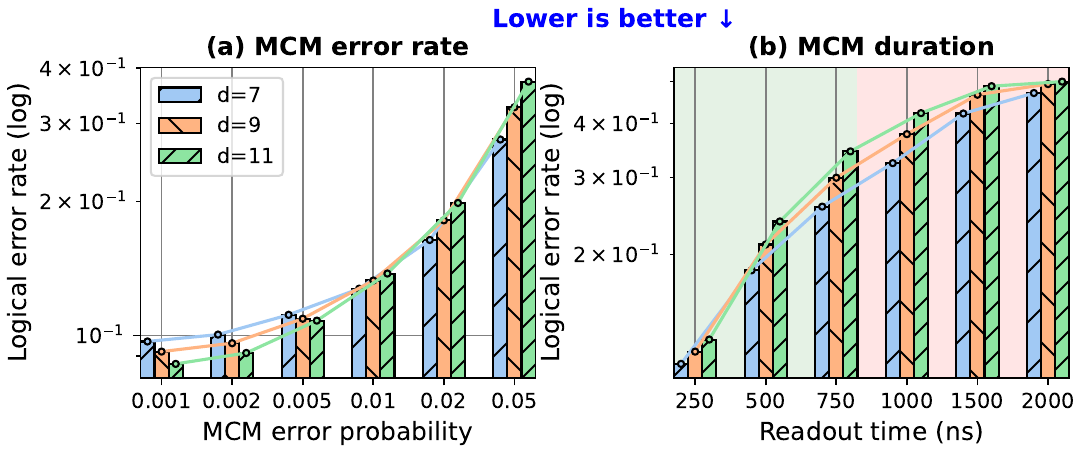}
   \caption{
   Logical error rate of the surface code on the Google Sycamore architecture \cite{Arute2019}. 
   {\em (a) Increasing with the MCM error rate. 
   (b) Increasing with the MCM duration. Green marks durations available with MCMit; red marks those available with HERQULES \cite{maurya2023scaling}.}
   }
   \label{fig:mcm_motivation}
\end{figure}

\myparagraph{Experimental setup}
{We simulate a fault-tolerant lattice-surgery merge within a lattice of surface code of distance 7, 9, and 11. After the merge, we perform $d+1$ rounds of stabilizer measurements. We simulate using the framework of Maurya et al. \cite{Maurya_2025} with a \textit{circuit-level noise model} derived from Google Sycamore \cite{Arute2019}. Syndrome measurements are decoded using a minimum-weight perfect matching (MWPM) decoder \cite{pymatching_higgott_2025}, with the logical error rate defined as the fraction of unsuccessful corrections over 500000 trials.}

\myparagraph{Problem \#1: MCM error effect on logical error rates}
{Fig.~\ref{fig:mcm_motivation} (a) shows the logical error rate change with the MCM error rate. We observe that reducing the MCM error by $10\times$ improves the logical error rate by 47.68\%, with the effect becoming more pronounced at higher measurement error probabilities.}

\myparagraph{Problem \#2: MCM latency effect on logical error rates}
{Fig.~\ref{fig:mcm_motivation} (b) shows how MCM duration affects the logical error rate during long execution. Compared to the 250ns duration, the standard 1000ns duration results in a logical error rate that is on average 212.98\% higher, while the 2000ns duration is 308.78\% higher. Interestingly, \projecttitle{} enables high-accuracy readout for short readout traces ($<$750ns), in contrast to the state-of-the-art HERQULES \cite{maurya2023scaling}.}

\takeaway{MCM errors and latency reduce the effectiveness of QEC: MCM errors directly increase the probability of correction failure, while longer readout durations amplify decoherence effects.}

\section{\projecttitle{} Overview}
\label{overview}

To address the MCM bottlenecks characterized in \S~\ref{motivation}, we propose \projecttitle{}, a hardware-software co-design comprising three mutually reinforcing layers: real-time multi-qubit branching, fast and high-accuracy qubit-state discrimination, and circuit-level optimization of dynamic circuits.

\begin{figure} [t]
   \centering
    \includegraphics[width=\columnwidth]{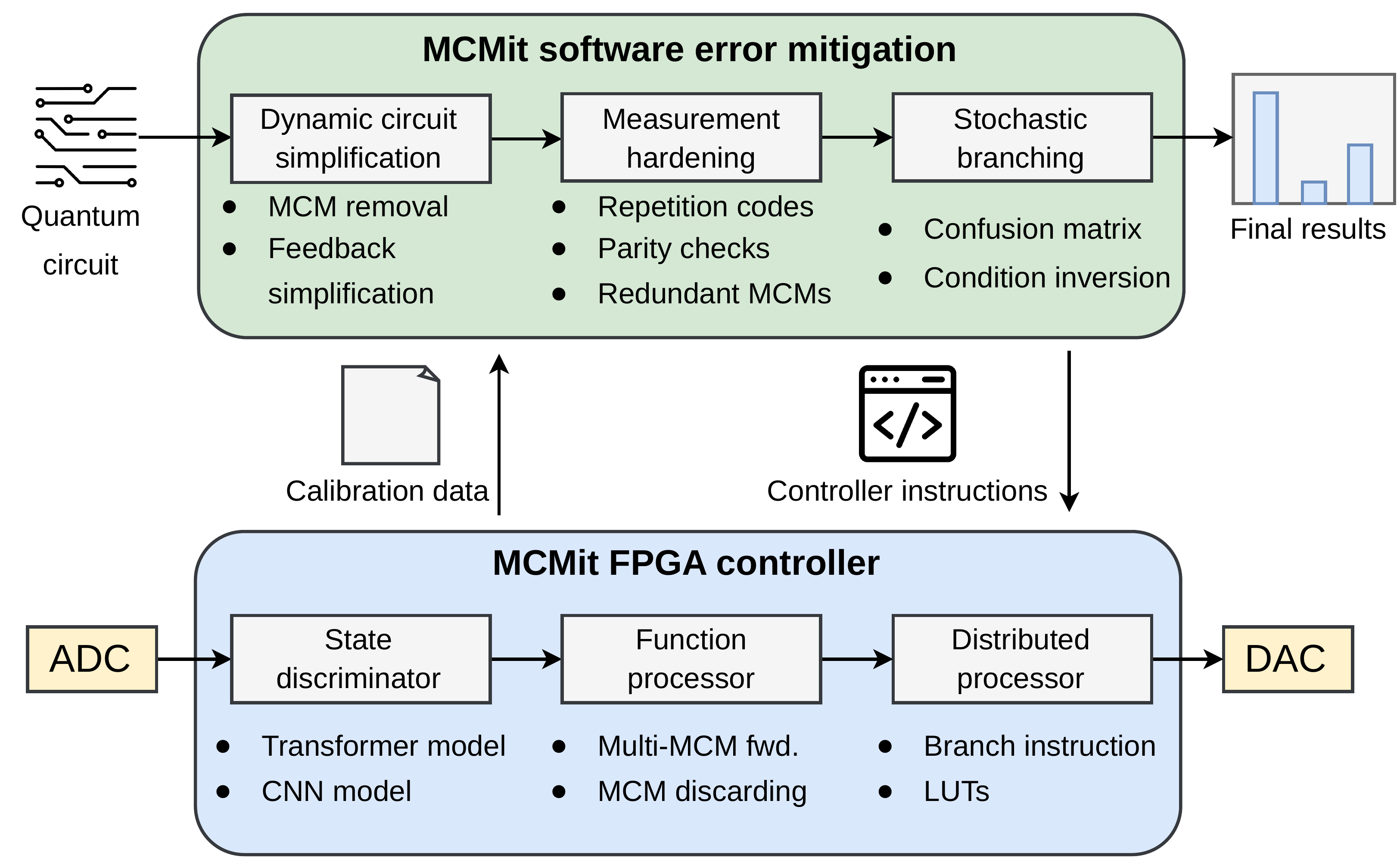}
   \caption{\projecttitle{} overview (\S~\ref{overview}). \textit{
    \projecttitle{} comprises the error mitigation software and the FPGA controller. 
    }}
    \label{fig:overview}
\end{figure}

\subsection{\projecttitle{} Architecture}
\label{overview:architecture}

Fig. \ref{fig:overview} shows the \projecttitle{} architecture, comprising the \projecttitle{} software layer for circuit-level error mitigation and the \projecttitle{} FPGA controller, which integrates the fast multi-qubit branching logic and the high-accuracy qubit-state discriminator model.

\myparagraph{Fast classical feedback}
\projecttitle{} introduces a fast multi-qubit branch instruction, which executes scalable conditional logic based on the joint outcomes of multiple qubits. This instruction enables more complex, constant-latency feedback protocols that are not possible with existing branching.
This functionality is particularly advantageous for routines such as parity checks and majority voting \cite{baumer2024efficient, xu2026distilling}, where MCM results are used to directly address a pre-computed LUT of correction operations, guaranteeing low-latency execution. 

\myparagraph{High-accuracy qubit-state discrimination}
We present two novel neural network architectures for qubit-state discrimination that operate directly on \textit{raw} measurement traces, eliminating the need for hand-crafted feature extraction or pre-classifiers. The first is a transformer that leverages self-attention mechanisms to learn complex temporal correlations within the trace, while the second is a lightweight residual CNN that exploits the local temporal correlations inherent in I/Q traces. These models achieve high accuracy even for \textit{short} readout traces, thereby enabling substantially faster measurement cycles that directly translate to lower logical error rates.

\myparagraph{Software error mitigation}
The \projecttitle{} software mitigates MCM-induced errors through three stages: \textit{dynamic circuit simplification} tracks qubit-state evolution to identify MCMs that can be replaced with probabilistic equivalents, and simplifies the corresponding classical control logic; \textit{measurement hardening} adds redundancy via repetition codes, parity checks, and repeated measurements; and \textit{stochastic branching} probabilistically modifies feedback conditions using hardware-provided confusion matrices to suppress branching errors.

\subsection{\projecttitle{} Workflow}
\label{overivew:workflow}

\begin{figure}
    \centering
    \includegraphics[width=\columnwidth]{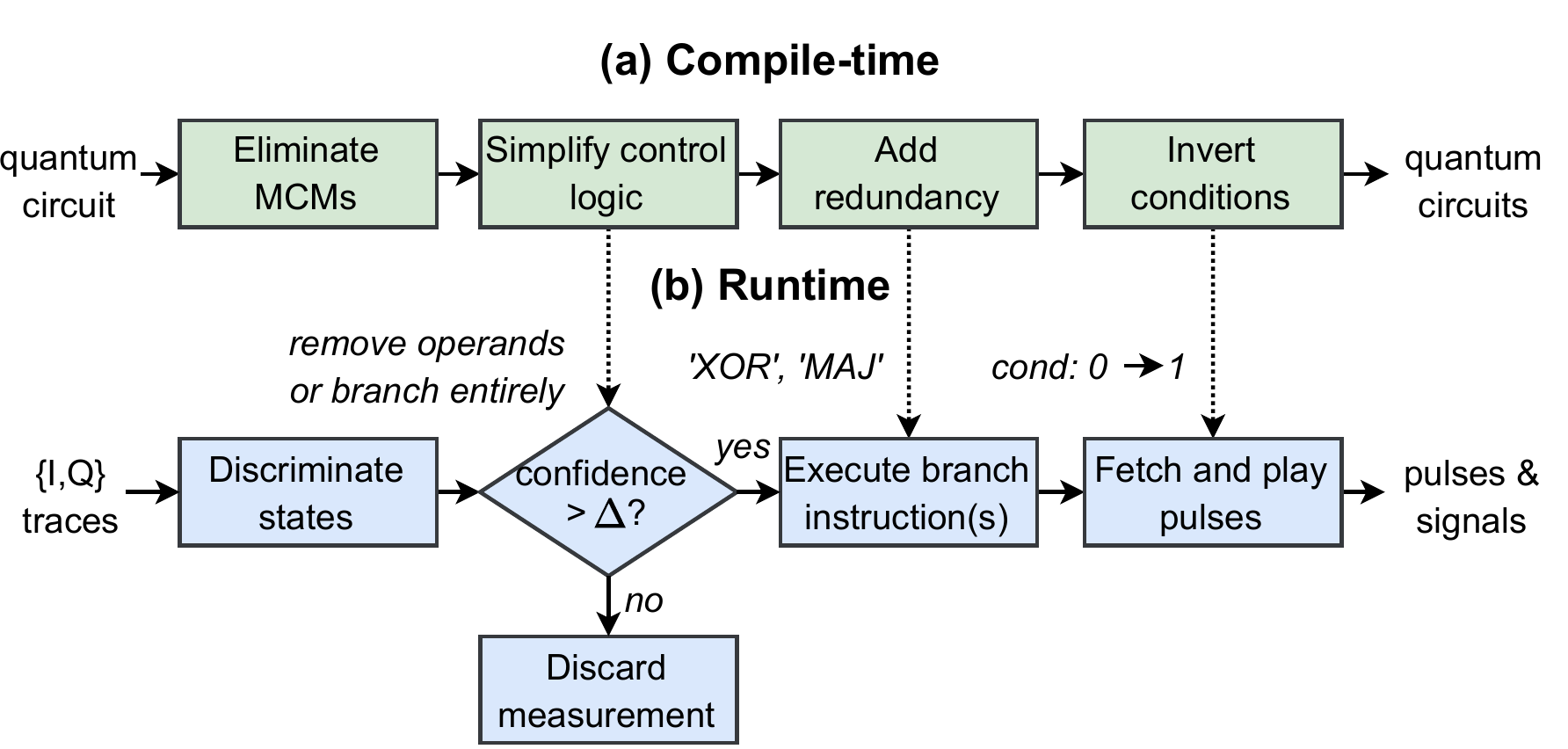}
    \caption{\projecttitle{} workflow (\S~\ref{overivew:workflow}).
    \textit{
    \textbf{(a)} Compile-time workflow and \textbf{(b)} runtime workflow.
    }    
    }
    \label{fig:workflow}
\end{figure}

Fig. \ref{fig:workflow} shows the compile-time and runtime workflows of \projecttitle{}. The software layer operates only at compile-time, while the FPGA controller layer operates only at runtime.

\myparagraph{Compile-time workflow}
To apply the first stage to the input quantum circuit, dynamic circuit simplification, we execute three sub-stages: (1) we identify and remove up to $M$ MCMs, where $M$ dictates the computational complexity of our static analysis tool, and (2), we simplify the control logic that is based on those MCMs. This removes some operands (e.g., XOR) in subsequent branch instructions or eliminates the branches entirely, effectively reducing classical latency. (3) We apply the measurement hardening pass, which adds redundancy via repetition codes, parity checks, and/or repeated MCMs. This stage modifies branch instructions by swapping ``equal'' for XOR or majority voting conditions on MCM results. (4) The stochastic branching stage inverts the branch conditions stochastically based on the bitflip error probabilities specified in the confusion matrices. 

\begin{figure*}[ht]
    \centering
    \includegraphics[width=0.9\textwidth]{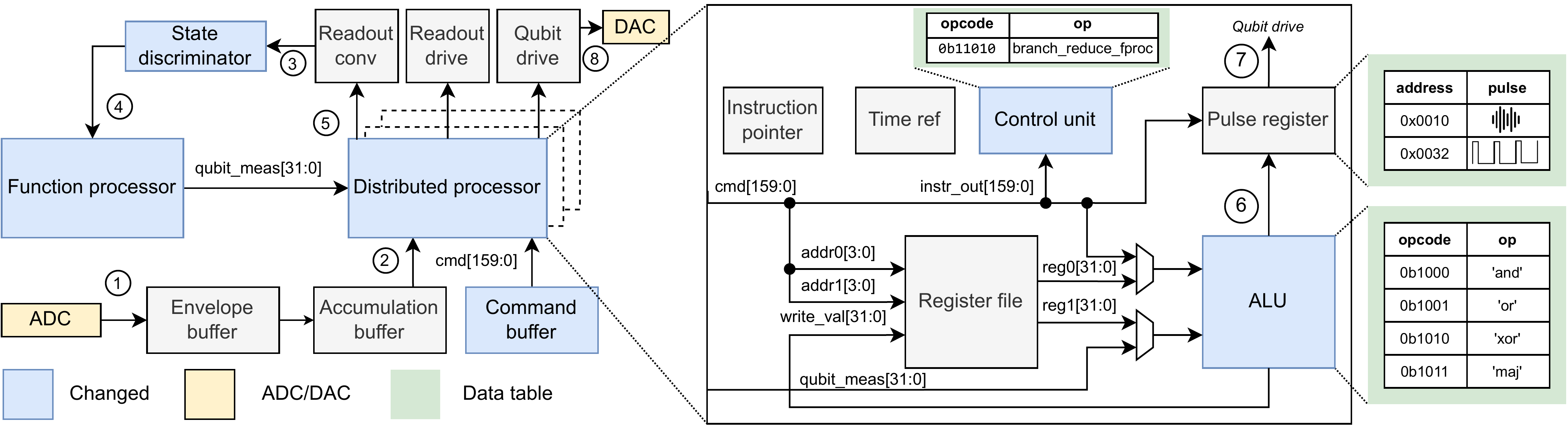}
    \caption{\projecttitle{} controller (\S~\ref{controller}).
    \textit{
    Blue boxes show modified components, yellow boxes show ADC/DAC components, and green boxes show example data tables. Steps (1)-(8) are executed for a conditional feedback operation based on an MCM result.
    }
    \vspace{-5pt}
    }
    \label{fig:controller}
\end{figure*}

\myparagraph{Runtime workflow}
For each set of MCMs that generates a set of \{I, Q\} traces, (1) we first discriminate the qubit-states and extract the discrimination confidence. (2) If it is below a threshold $\Delta$, we discard (some of) the measurements, and the shot must be rerun. (3) We execute the branch instruction(s) that execute the feedback operation. The instruction's operands and conditions are already modified by the software layers as previously discussed. (4) We fetch the pulses to play from memory, if any, and we send them to the qubit drive, where the pulses eventually become signals at the DAC.

\section{Enabling Scalable Classical Feedback}
\label{controller} 

Existing controllers only support high-latency, single-qubit branching that cannot natively execute the complex multi-qubit conditional logic required by DQC and QEC (\S~\ref{motivation}). We address this by introducing a controller with a constant-latency multi-input branch instruction that natively supports multi-qubit operations.

\subsection{\projecttitle{} Controller Overview}
Fig. \ref{fig:controller} shows the design of our FPGA controller, based on the open-source Qubic framework \cite{xu2021qubic}. At a high level, the controller comprises a \textit{function processor}, a \textit{distributed processor}, a set of \textit{buffers}, and a set of \textit{interfaces}. 

\myparagraph{Function processor}
The function processor is responsible for applying processing functions to qubit readout results and, as such, supports an interface between external components, such as state discriminators or dedicated QEC decoders, if needed. 

\myparagraph{Distributed processor}
The distributed processor comprises multiple cores, each managing interfaces to the function processor and inter-core synchronization. Cores execute programs featuring precisely timed pulse commands, controlling AWG characteristics like frequency and envelope shape. Concurrently, cores implement arbitrary control flow, such as conditional branching, based on external inputs or internal state using a 16x 32-bit register bank and an ALU that performs integer comparisons and arithmetic. ALU results can update registers, adjust the timing reference, or modify the program counter, enabling dynamic execution paths. 
The distributed processor natively implements branching logic using LUTs, processing MCMs results in parallel as they are dispatched from the function processor, enabling simultaneous classical execution.

\myparagraph{Buffers}
There are four types of buffers: (1) the \textit{envelope} buffer, which stores the real and imaginary parts of the pulses' envelopes, (2) the \textit{accumulation} buffer, which stores the processed {I/Q} trace after down-converting and filtering, (3) the \textit{command} buffer, which stores the detailed information for every command, and (4), the \textit{acquisition} buffer, which we omit for simplicity in the context of this work.

\myparagraph{Interfaces}
There are three types of interfaces: (1) the \textit{readout converter}, which performs the final trace conversion to a format compatible for state discrimination, (2) the \textit{readout drive}, which sends readout pulses through the DACs, and (3) the \textit{qubit drive}, which sends gate pulses through the DACs.

\subsection{Multi-qubit Branch Instruction}
\label{branch_instruction}

To enable an efficient and scalable multi-qubit branch instruction, we modify the architecture described above, as shown in Fig. \ref{fig:controller}, and highlight the modifications in blue.

\myparagraph{ALU}
We modify the ALU with a wider opcode to support new operations, specifically \texttt{'and},  \texttt{'or},  \texttt{'xor}, and  \texttt{'maj}, which performs majority voting. All new operations support up to 32 qubit inputs and 32-bit masks to select them. Specifically, the ALU receives a 32-bit mask that selects the qubit measurements to operate on as its second input.

\myparagraph{Control unit}
We modify the control unit to support a wider ALU opcode, and add our new branch instruction \texttt{branch\_reduce\_fproc} (Fig. \ref{fig:controller} (top right)).

\myparagraph{Command buffer}
We modify the command buffer by extending the command length since the ALU requires new opcodes for operations on multiple qubits. As the command memory is organized into 32-bit blocks, we add a new block, increasing the command memory and bus width from 128 bits to 160 bits (5 x 32 bus).

\myparagraph{Function processor}
Originally, during the single-qubit branch instruction, the distributed processor core requests the qubit measurement from the function processor and receives it as the least significant bit of a 32-bit bus. We modify the function processor to send the whole 32-bit measurement register containing all qubit measurements to the ALU in the case of \texttt{branch\_reduce\_fproc}.

\myparagraph{State discriminator}
Lastly, we modify the state discriminator by implementing a transformer and a CNN, which we explain in \S~\ref{state-discrimination}.

\subsection{Controller Workflow}

We enumerate the steps the controller will take during an MCM and a dependent feedback operation, as shown in Fig. \ref{fig:controller}. We assume that previously the readout drive sent a readout pulse; thus, we start from (1), the gathering of the ADC data, which are accumulated in the accumulation buffer. These data are passed to the distributed processor (2), then to the readout converter, which produces the final {I, Q} trace (3). The state discriminator outputs a 1 bit measurement result per qubit and sends it to the function processor (4), which, in this case, simply forwards all qubit measurements to the distributed processor (5). Here, the control unit and the ALU fetch the \texttt{branch\_reduce\_fproc} with the appropriate ALU op (e.g., \texttt{'maj'}) (6). When the branch is decided, the branch's pulse is fetched from the memory using the pulse register (7), and the pulse is sent to the DAC by the qubit drive (8).

\begin{figure} [t]
    \centering
    \includegraphics[width=0.9\columnwidth]{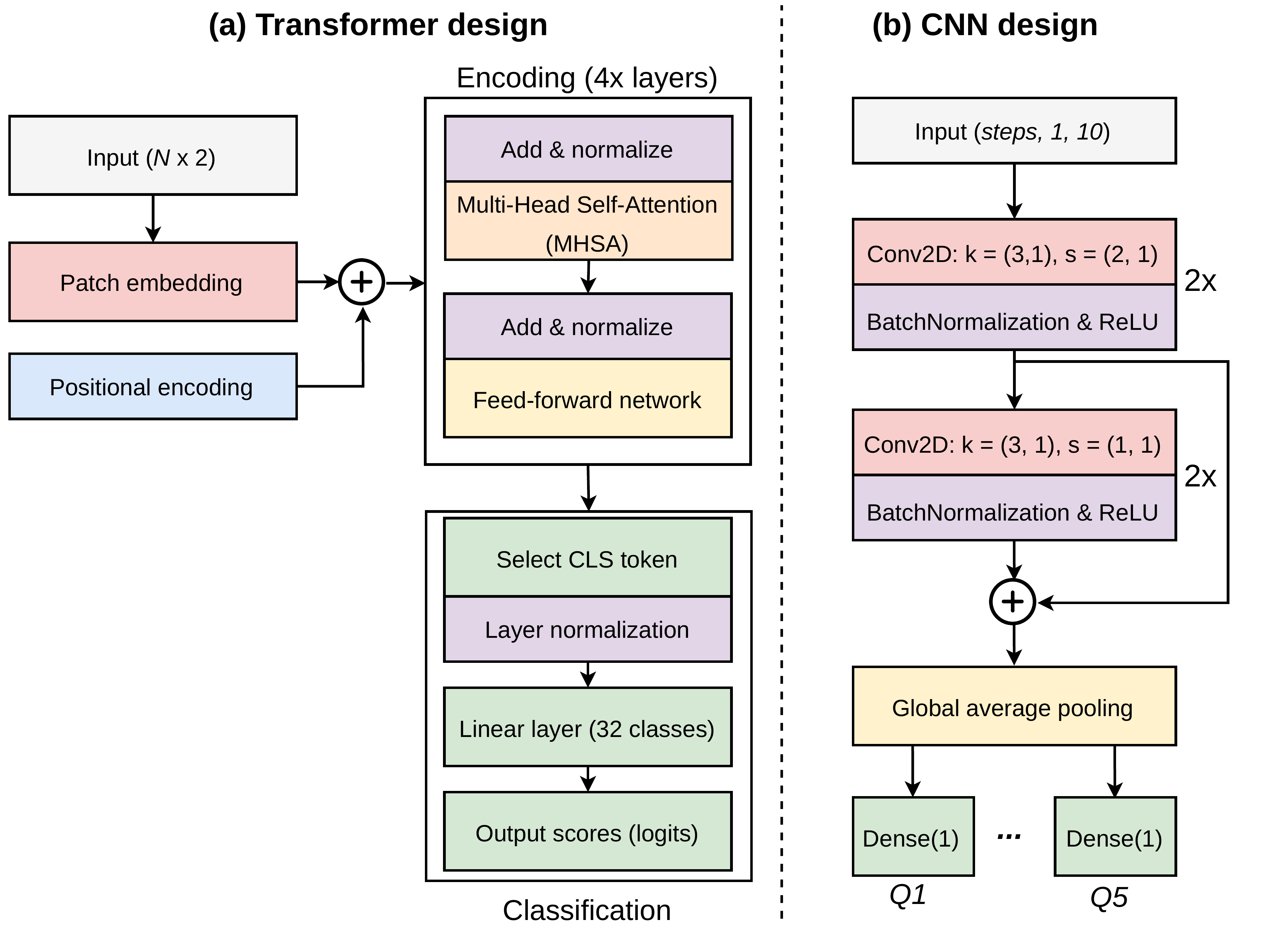}
    \caption{The \projecttitle{} qubit-state discriminators (\S~\ref{state-discrimination}).}
    \label{fig:fnn-transformer}
\end{figure}

\section{High-Accuracy Qubit-State Discrimination}
\label{state-discrimination}

Fast readout and high-accuracy single-shot qubit-state discrimination lies at the foundation of the DQC and QEC paradigms (\S~\ref{motivation}). Any enhancement in the readout speed (i.e., shortening the readout length) or discrimination accuracy directly improves the overall fidelity of the entire quantum computation. In this section, we introduce two discriminator designs that operate directly on raw I/Q measurement traces: a transformer-based model that leverages self-attention to capture global temporal dependencies (\S~\ref{sec:transformer}), and a lightweight residual CNN that exploits local temporal correlations while remaining parameter-efficient (\S~\ref{sec:cnn}).

\myparagraph{Limitations of existing approaches}
Previous work on qubit-state discrimination has explored Match Filters (MFs), Support Vector Machines, and FNNs \cite{lienhard2022deep}. HERQULES \cite{maurya2023scaling} proposed a lightweight FNN combined with MFs to balance accuracy and hardware constraints. Unfortunately, these approaches share a fundamental limitation: they either aggregate the entire readout trace into a single feature vector via boxcar integration (MFs, FNNs) or operate on hand-crafted summary statistics, discarding the temporal structure of the signal entirely. As a result, such architectures are \textit{not suitable for short readout traces}, as we show in Table \ref{tab:model_accuracies_short}.
By contrast, a transformer's multi-head self-attention captures non-local temporal correlations, attending to the most informative segments of a noisy pulse even under minimal integration time, while a CNN's strided convolutional filters efficiently exploit local temporal structure directly on raw I/Q traces, without hand-crafted feature extraction or pre-classifiers, making both architectures inherently more suitable for the short-trace, low-SNR regime.

\subsection{Transformer Design}
\label{sec:transformer}

\myparagraph{Design rationale}
Before committing to a lightweight architecture, we deliberately explore an expressive, parameter-rich model to establish an accuracy ceiling for the readout discrimination task. The key question is whether a model with hundreds of thousands of parameters and global self-attention can extract discriminative signal that smaller networks fundamentally miss, particularly at short readout durations where the SNR is low. If the transformer significantly surpassed existing discriminators such as HERQULES, this would indicate that long-range temporal correlations carry information that no current lightweight model captures, warranting further investigation into knowledge distillation or architectural insights. Conversely, if a subsequent lightweight design matches the transformer, we can be confident that the deployable model is not leaving accuracy on the table. As we show in \S~\ref{evaluation:discrimination}, the latter turns out to be the case: our CNN matches the transformer at a fraction of the FPGA cost, validating it as the right design point.

\myparagraph{Transformer architecture}
Fig.~\ref{fig:fnn-transformer} (a) shows the transformer encoder-only architecture.
The transformer operates on input traces of shape $(N, 2)$ (by default $N=500$). The trace is partitioned into non-overlapping temporal patches (10 samples each), which are linearly projected into an embedding space and prepended with a learnable [CLS] token that aggregates sequence-wide information. Positional encodings are added to preserve timing information, and the sequence is processed by four encoder layers of multi-head self-attention to capture long-range temporal dependencies. The final [CLS] state is fed to a feed-forward classifier to produce the qubit-state classification logits. We train the model using AdamW with weight decay and a Cosine Annealing scheduler, with label smoothing and dropout for regularization. The output represents five qubits, i.e., 32 binary states.

\myparagraph{Limitation} Although the transformer achieves high qubit-state discrimination accuracy even for short readout traces (\S~\ref{evaluation:discrimination}), it also exhibits a large size that stresses FPGA resources (Table \ref{tab:fpga_utilization}). To this end, we propose a CNN design.

\subsection{CNN Design}
\label{sec:cnn}
We detail the architecture of the CNN discriminator (Fig. \ref{fig:fnn-transformer} (b)).

\myparagraph{Input pre-processing}
Because the raw multiplexed readout trace overlaps every qubit's response in both time and frequency, feeding it to the network unprocessed would waste model capacity on separating channels rather than discriminating states, so we isolate each qubit's channel before it reaches the CNN. To avoid depending on nominal hardware specifications, we calibrate this separation once per setup, picking for every qubit $i$ the demodulation frequency $f_i^{*} = \arg\max_{f} |S_1(f) - S_0(f)|$ that maximizes the spectral separation between its ground- and excited-state magnitude spectra. Using these fixed frequencies, each channel is then downconverted to baseband and low-pass filtered to suppress out-of-band noise and inter-channel interference, and decimated to shrink the memory footprint, before being stacked into a tensor $X \in \mathbb{R}^{T\times N \times 2}$ across $N$ qubits' I/Q components. We evaluated several filter and decimation configurations for this pipeline and found accuracy relatively insensitive to the exact choice, so we treat them as tunable rather than fixed parameters.

\begin{figure*}[t]
    \centering
    \includegraphics[width=0.9\textwidth]{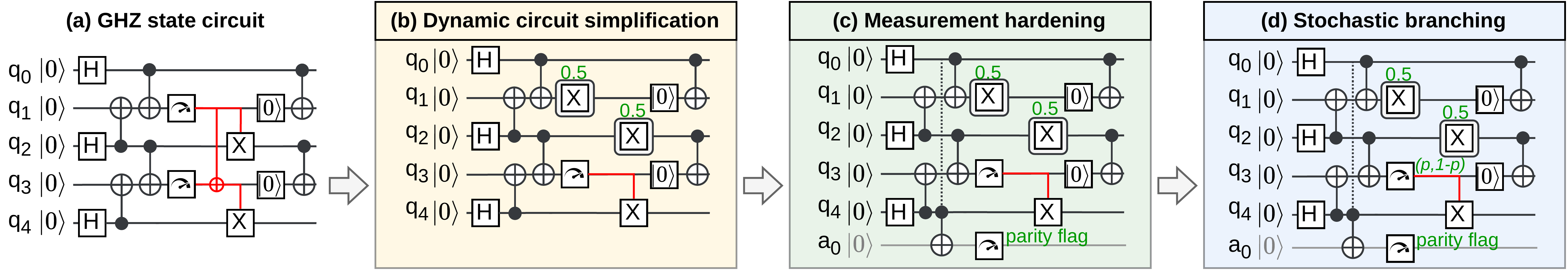}
    \caption{Software MCM error mitigation (\S~\ref{compiler}). 
        \textit{
            \textbf{(a)} GHZ state dynamic circuit. 
            \textbf{(b)} Dynamic circuit simplification removes MCMs when possible and simplifies classical conditional logic. \textbf{(c)}  Measurement hardening leverages repetition codes, parity checks, and flag qubits to detect and/or correct errors. \textbf{(d)} Stochastic branching factors MCM errors into branching decisions.
        }
    }
    \label{fig:compiler_overview}
\end{figure*}

\myparagraph{Architecture}
Because the FPGA controller operates under a strict latency and resource budget (\S~\ref{evaluation:discrimination}), we favor a compact, largely convolutional design over deeper or fully-connected alternatives; across the several such lightweight variants we evaluated, differences in discrimination accuracy were small, so we describe the configuration used throughout our evaluation (Fig.~\ref{fig:fnn-transformer}~(b)). The preprocessed tensor is first reshaped into a multi-channel representation, with each qubit's I/Q components mapped to its own channel, and passed through an input section of two strided convolutional layers (kernel size 3, stride 2); replacing pooling with strided, learnable downsampling lets the network decide how to compress the sequence rather than discarding signal nuances arbitrarily, while the small kernels keep the parameter count low. Since the upstream signal processing has already surfaced much of the discriminative structure, a single residual block is enough to refine it further, learning incremental corrections that capture subtle cross-channel effects such as measurement crosstalk without distorting the dominant per-qubit features. To keep the model compact and robust to temporal shifts in the readout pulse rather than relying on parameter-heavy fully connected layers over flattened features, global average pooling then collapses the temporal dimension into a fixed-length vector. Finally, so that the output scales linearly rather than exponentially with the number of qubits while still implicitly capturing inter-qubit correlations through the shared convolutional trunk, five independent sigmoid neurons (one per qubit) produce binary predictions under a multi-label formulation.

\myparagraph{Training and inference}
To let the network converge quickly while still fine-tuning in later epochs, we train with the Adam optimizer under a step-decay schedule (initial learning rate $5\times10^{-4}$, decayed by a factor of $0.95$ every 3 epochs); since each qubit's state is only weakly coupled to the others, we supervise each output independently with binary cross-entropy loss rather than treating the 32 joint states as a single classification target. We use a 65/35 train/validation split, a batch size of 64 over 30 epochs, and track per-qubit accuracy and the geometric-mean fidelity across qubits, though we found these training hyperparameters had only a mild effect on the final accuracy across the configurations we tried. Because the FPGA controller cannot host a full-precision model within its latency and resource budget, we apply post-training INT8 quantization after training (Table~\ref{tab:fpga_utilization}), which costs negligible accuracy given the already-compact dynamic range of the preprocessed input. For inference, a single raw downsampled trace is fed directly into the model, and the five sigmoid outputs are thresholded at 0.5 to yield the per-qubit state predictions.

\begin{figure}[t]
    \centering
    \includegraphics[width=0.85\linewidth]{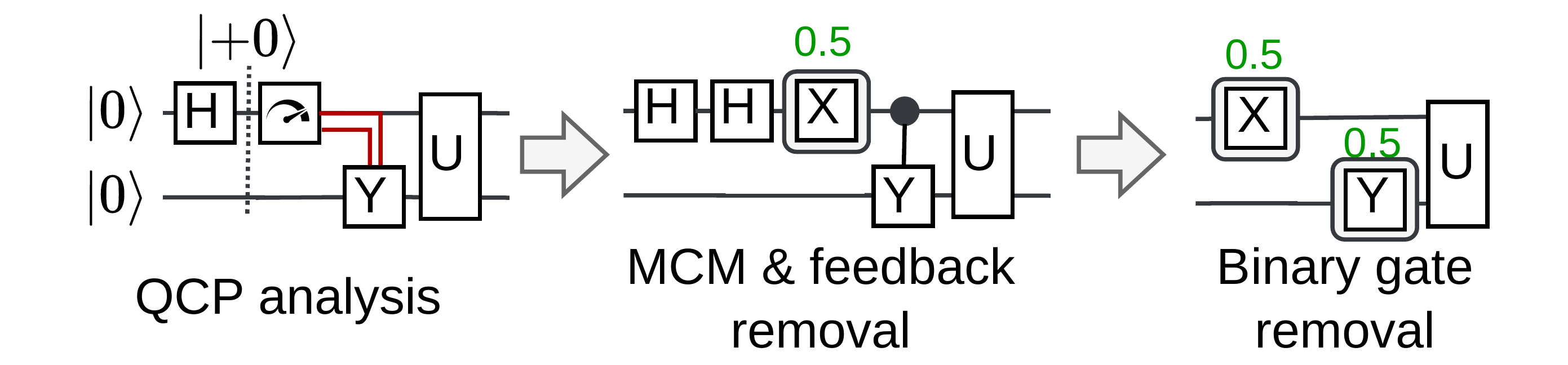}
    \caption{Dynamic circuit simplification (\S~\ref{compiler:simplification}). 
    \textit{The original dynamic circuit is replaced by a static sub-circuit containing a probabilistic gate, whose outcome is decided at compile time.}}
    \label{fig:MCM_QCP_illustration}
\end{figure}

\section{Software MCM Error Mitigation }
\label{compiler}

To mitigate MCM-induced branching errors that propagate through the circuit (\S~\ref{motivation}), we present a three-stage pipeline: (1) static MCM elimination via symbolic reasoning to remove measurements and their associated control logic (\S~\ref{compiler:simplification}), (2) measurement hardening via repetition codes, parity checks, and repeated measurements to detect and correct bitflip errors (\S~\ref{compiler:QEC}), and (3) stochastic branching, which probabilistically modifies control flow conditions using hardware-provided bitflip probabilities to reduce incorrect branching (\S~\ref{compiler:stochastic_branching}).

\subsection{Dynamic Circuit Simplification} 
\label{compiler:simplification}
The first stage simplifies dynamic circuits by removing MCMs when possible, and then simplifies or eliminates their dependent feedback operations. To remove MCMs, we track the qubit states from initialization until the MCM, without using full quantum simulation, based on the quantum constant propagation technique \cite{chen2023qcp}.{Importantly, this technique is generally not applicable in QEC settings, where syndrome MCMs are crucial for decoding and subsequent corrections. However, it is applicable in any non-FTQC case.}

\myparagraph{Quantum constant propagation (QCP)}
QCP is introduced as a symbolic execution technique that improves circuit optimization by identifying and removing redundant gates based on static information available at compile time \cite{chen2023qcp}.
QCP tracks quantum states using entanglement-aware symbolic representations, grouping entangled qubits via a union-based data structure. To ensure scalability, QCP imposes a configurable upper bound on the size of tracked entangled sets (parameterized by $N_{max}$). When this threshold is exceeded, QCP conservatively abandons tracking and assigns the unknown value $\top$. As such, QCP avoids simulation overheads and maintains polynomial runtime complexity.

\myparagraph{MCM removal}
The work by Chen et al. \cite{chen2024mcm, chen2025mcm} extends QCP to support circuits with MCMs, demonstrating that many MCMs can be removed during compilation by treating them as static observations under certain conditions. The core idea is shown in Fig. \ref{fig:MCM_QCP_illustration}: in specific sub-circuits, qubit values at measurement points can be inferred symbolically from initial states and prior gates. When symbolic execution reveals a deterministic value at the point of an MCM, the measurement and its associated classical feedback logic can be replaced by a probabilistic sub-circuit. Naturally, this improves execution fidelity and reduces latency by eliminating unnecessary measurement and feedback operations.

\myparagraph{Limitations and our solutions}
This previous work is characterized by two main limitations, which we address in our work: (1) a lack of multi-qubit classical control, since they assume branching based on a single MCM result, and (2) unnecessary 2q-gate overhead, where classical control is replaced with noisy and slow 2q-gates (Fig. \ref{fig:MCM_QCP_illustration}, middle). For (1), we extend symbolic reasoning to support multi-qubit classical control logic, allowing the compiler to simplify complex feedback operations such as XOR or majority-vote outcomes, effectively relaxing FPGA hardware requirements and lowering classical feedback latency. For (2), we further eliminate 2q gates and replace them with probabilistic 1q-gates. This leads to removing the MCM, the associated classical feedback, and the 2-qubit gates entirely.

\myparagraph{Example}
Fig. \ref{fig:compiler_overview} (b) shows the dynamic circuit simplification pass. The first MCM (on qubit $q_1$) and its dependent feedback $X$ operation are replaced with two probabilistic $X$ gates, with 0.5 probability each. As such, in every shot compilation, one out of four possible circuits will be compiled. Moreover, the propagation of the MCM result to $q_3$ and the XOR operator are removed entirely.

\subsection{Measurement Hardening}
\label{compiler:QEC}

To mitigate MCM errors, we adopt lightweight primitives inspired by QEC, including repetition codes, parity checks, and repeated measurements with majority voting. These techniques serve three purposes: (1) They reduce the likelihood of multiple simultaneous MCM errors through redundancy, (2) enable error detection by exposing bitflips of correlated measurements, and, in the case of repetition codes and repeated measurements, (3) allow for error correction. 

\myparagraph{Repetition codes}
Repetition encoding detects bitflip errors in MCMs by redundantly encoding a logical qubit into $d$ physical qubits and comparing measurement outcomes \cite{gunther2021improving, wootton2018repetition}. Even $d$ allows error detection via parity checks, while odd $d$ enables single‑bitflip correction by majority voting. Since CNOT gates used for encoding/decoding also introduce errors, these must be considered. As such, the resulting error probability for a distance‑$d$ code with $N$ CNOTs and $d$ measurements is $p^{rep}_{err} = 1 -(1-p_{CNOT})^N(1-p_{meas})^d$, where $p_{CNOT}$ and $p_{meas}$ are the per‑gate and per‑measurement error probabilities, respectively. All encoded qubits are measured simultaneously, and the final outcome is determined using a two-qubit logical AND (for detection) or a three-qubit majority vote (for correction). Notably, these operations are natively supported by the \projecttitle{} controller at a constant latency cost (\S~\ref{branch_instruction}).

\begin{table}[t]
\centering
\footnotesize
\caption{Qubic's latency overhead compared to \projecttitle{} for $N$-qubit conditional operations. Classical feedback includes {DAC, ADC, pulse preparation, and state discrimination.}}
\label{tab:branch_latency}
\label{tab:latency_full_comparison}
\begin{tabular}{lcc}
\toprule
\multirow{2}{*}{\textbf{N-qubit input (Operation)}} & \multicolumn{2}{c}{QubiC latency (ns)}  \\
\cmidrule(lr){2-3}
 & Branch instr. &  Classical feedback \\
\midrule
N=1  (Single Branch) & 16 $(\mathbf{\times1})$  & {205} $(\mathbf{+0\%})$ \\
N=7  (XOR/Majority) & 112 $(\mathbf{\times7})$ & {301} $(\mathbf{+46\%})$  \\
N=8  (XOR) & 128 $(\mathbf{\times8})$ & {317} $(\mathbf{+54\%})$  \\
N=15 (XOR/Majority) & 240 $(\mathbf{\times15})$ & {429} $(\mathbf{+109\%})$ \\
N=16 (XOR) & 256 $(\mathbf{\times16})$ & {445} $(\mathbf{+117\%})$ \\
{N=32 (XOR)} & {512 $(\mathbf{\times32})$} & {701} $(\mathbf{+241\%})$ \\
\hdashline
{N=128 (XOR)} & {2048 $(\mathbf{\times128})$} & {2237} $(\mathbf{+991\%})$ \\
\bottomrule
\end{tabular}
\end{table}

\myparagraph{Parity checks} To protect long-range GHZ states \cite{baumer2024efficient}, we use stabilizer-based parity checks that detect MCM-induced bitflip errors. For an $n$-qubit GHZ state $\ket{GHZ_n}$ the stabilizer group is generated by $\braket{X_1X_2...X_n, Z_1Z_2, ..., Z_{n-1}Z_n}$, and a subset of the $Z_iZ_j$ stabilizers is non‑destructively measured via the ancillae, yielding outcomes $m_k\in{+1,-1}$ for each $P_k=Z_{i_k}Z_{j_k}$. Any $m_k=-1$ signals a parity violation. 
 We define the error-detection flag $D=1-\prod_{k=1}^{r} \frac{1+m_k}{2}$, such that $D=1$ whenever a bitflip is detected, allowing faulty shots to be discarded and improving overall fidelity without requiring full error correction.

\myparagraph{Example}
Fig. \ref{fig:compiler_overview} (c) shows the measurement hardening stage, specifically a parity check. Here, we add an ancilla qubit ($a_0$) and entangle it with qubits $q_0, q_2, q_4$ to perform a stabilizer-based parity check.

\subsection{Stochastic Branching}
\label{compiler:stochastic_branching}

To address branching errors caused by MCM bitflips in dynamic circuits, we leverage confusion matrix \cite{nation2021scalable} data provided by the \projecttitle{} controller to inform stochastic compilation of control flow. Specifically, for measurement-conditioned operations, the compiler uses per-qubit bitflip probabilities to probabilistically invert the branching condition across repeated shot executions. For example, if a feedback gate is conditioned on a measured `1' with bitflip probability $p$, the compiler \textit{partitions} the existing $N$ shots: approximately $(1-p) \cdot N$ instances run with the original condition and $p \cdot N$ with the condition flipped, mitigating control flow divergence across shots. Crucially, the total shot count remains $N$; no additional shots are introduced.

Formally, for $k$ independent branches, each shot independently samples its branch conditions
from the joint distribution. The dominant configuration (no flips) retains
$(1-p)^k \cdot N$ shots, while the most important minority configurations (single flips)
each receive $p \cdot (1-p)^{k-1} \cdot N$ shots. Adequate statistical coverage requires
$N \geq S_{\min} / (p \cdot (1-p)^{k-1})$, where $S_{\min}$ is the minimum number of shots
per configuration. For small $p$ this simplifies to $N \gtrsim S_{\min}/p$: with $p=0.03$
and $S_{\min}=30$, approximately $N \geq 1{,}000$ shots suffice, well within standard
execution budgets (we use $N=10{,}000$). Higher-order configurations ($j \geq 2$
simultaneous flips) occur with probability $\binom{k}{j} p^j (1-p)^{k-j}$, dropping
rapidly and contributing negligibly to overall fidelity.

\myparagraph{Example}
Fig. \ref{fig:compiler_overview} (d) shows qubit $q_4$ which is unprotected from branching errors, since the MCM at qubit $q_3$ can be erroneous. We compile the circuit as-is $p \cdot N$ times, and $(p-1)\cdot N$ times without the $X$ gate, where $p$ is the readout error probability for this qubit.

\subsection{Applicability Across Application Classes}
The three software stages target complementary application scenarios.
Dynamic circuit simplification (\S~\ref{compiler:simplification}) applies to DQC and
non-FTQC workloads, where MCMs can sometimes be statically resolved,
but is generally not applicable in QEC settings where syndrome MCMs
are essential for decoding. In contrast, measurement hardening
(\S~\ref{compiler:QEC}) and stochastic branching (\S~\ref{compiler:stochastic_branching})
apply broadly across all application classes, including QEC, as they
mitigate errors in MCMs that must be retained. As a result, the relative contribution of each stage varies with the circuit's MCM structure, e.g.,
circuits with many eliminable MCMs (e.g., constant-depth GHZ) benefit
most from simplification.

\begin{figure}[t]
    \centering
        
    \includegraphics[width=\columnwidth]{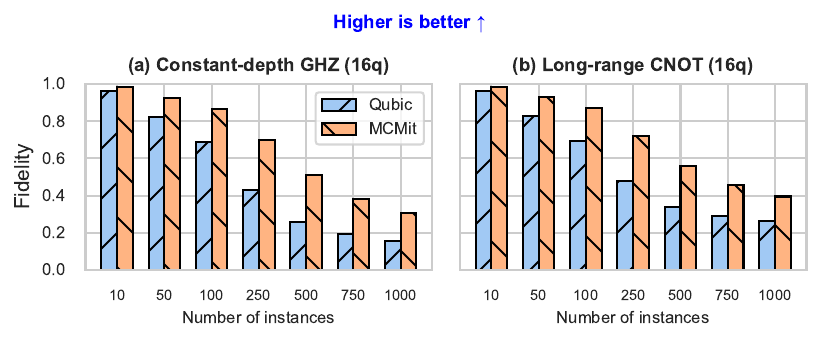}
    
    \caption{Classical feedback latency impact (\S~\ref{evaluation:classical_feedback}).
    \textit{
    The x-axis shows the number of instances of the GHZ/CNOT circuit in a higher-level application. \projecttitle{} achieves $57.3\%$ and $37.8\%$ higher fidelity than Qubic, on average.
    }    
    }
    \label{fig:feedback_impact}
\end{figure}

\section{Evaluation}
\label{evaluation}

\subsection{Experimental Methodology}

\myparagraph{Implementation}
We implement the \projecttitle{} controller on top of Qubic 2.0 \cite{xu2021qubic} by modifying the Qubic software (Qubic IR, compiler, assembler, and command generation), the Qubic distributed processor, and the Qubic gateware. We implement the \projecttitle{} transformer and CNN with PyTorch v1.7.0., TorchVision v.0.8.1., and hls4ml v.1.2.0 \cite{hls4ml_concepts, fast_hls4ml}. Last, we implement the \projecttitle{} software error mitigation using Qiskit v.2.1.0 \cite{javadiabhari2024qiskit}, Qiskit Aer v.0.17.1, Networkx v.3.5., and Qiskit mthree v.3.0.0. 

\myparagraph{Quantum hardware}
We conduct experiments requiring quantum hardware on \texttt{ibm\_fez}, an IBM Heron r2 QPU \cite{ibm-devices}. However, since IBM \textbf{does not} give public access to readout traces, we use the traces of Lienhard et al. \cite{lienhard2022deep}. Their QPU comprises five frequency-tunable superconducting transmon qubits with alternating target frequencies, capacitive nearest-neighbor coupling, and dispersive coupling to a dedicated readout resonator (bare frequencies ~7 GHz); the measured qubit lifetimes ($T_1$) range from 12--41$\mu s$.

\myparagraph{Readout trace dataset}
The dataset comprises 1.6 million measurement traces experimentally extracted from \cite{lienhard2022deep}. Each trace consists of sequentially recorded \{I, Q\} values captured at a sampling rate of 500 MSamples/sec over a duration of 1$\mu s$, representing one of the $2^5=32$ possible computational basis states. The dataset includes 50,000 distinct traces for each of these 32 basis states.

\myparagraph{Classical hardware}
For classical tasks (such as ML training), we use a server with a 64-core Intel Xeon Gold 6326 CPU at 2.90GHz and 314GB RAM, and an A40 Nvidia GPU and 48GB VRAM. For our controller implementation, we use the Xilinx Alveo XCU280 data center acceleration card with 8GB HBM.

\myparagraph{Benchmarks}
We use four micro-benchmarks: \textbf{(1)} Constant-depth GHZ state and \textbf{(2)} Long-range CNOT (i.e., CNOT between non-local qubits), both using dynamic circuits as defined in \cite{baumer2024efficient}, \textbf{(3)} quantum teleportation in a ladder structure, and \textbf{(4)} repeated teleportation (on the same qubits) \cite{bennett1993teleporting}. The quantum state we teleport is $\ket{\psi} = \cos\frac{\pi}{8} \, e^{-i 3\pi / 8} \, \ket{0}- i \, \sin\frac{\pi}{8} \, e^{i 3\pi / 8} \, \ket{1}$, to ensure robustness against bitflip errors. Last, we use two high-level applications: MECH \cite{hezi2024MECH}, the SOTA compiler for chiplet architectures, and the surface QEC code \cite{horsman2012surface}, as described in \S~\ref{motivation:qec}.

\myparagraph{Metrics}
We distinguish the evaluation metrics based on a component basis: \textbf{(1)} For the \texttt{branch\_reduce\_fproc} instruction, we measure \textit{latency} (in ns), from when the branch opcode starts executing until the if/else pulse starts executing. 
\textbf{(2)} For the discriminators, we measure discrimination \textit{accuracy} ($[0,1]$), discrimination \textit{latency} (in ns), and FPGA resource utilization. 
\textbf{(3)} For the software error mitigation, we measure \textit{Hellinger fidelity} as defined in \cite{fidelity-qiskit}, where it ranges in $[0,1]$ and higher is better.

\myparagraph{Baselines}
We distinguish our baselines based on the \projecttitle{} component we evaluate. For the \texttt{branch\_reduce\_fproc} instruction, we use  Qubic \cite{xu2021qubic} as the baseline, since we do not have access to proprietary industrial FPGA controllers.
For the \projecttitle{} discriminators, we use the baseline FNN design by Lienhard et al. \cite{lienhard2022deep},{Qubic's ML discriminator \cite{vora2024ml}}, and the state-of-the-art HERQULES system \cite{maurya2023scaling}. Last, for the \projecttitle{} software error mitigation, we use Qiskit mthree \cite{nation2021scalable} as the baseline, since it is the SOTA readout error mitigation method consistently used by IBM \cite{javadiabhari2025big, baumer2024quantum, barron2024provable, motta2023quantum}.

\myparagraph{Training methodology}
For each readout length, we train \projecttitle{} from scratch on traces truncated to that length. For HERQULES, we instead follow its published ``without re-training'' protocol: a single network is trained on the full $1000$\,ns window and applied to shorter traces by truncating the integration window (matched-filter envelopes re-sliced, network weights frozen). Both systems are trained and evaluated on the same calibration dataset with a fixed train/test split, using a matched-filter preprocessing without cross-qubit shot mixing.

\begin{table}[t]
\centering
\footnotesize
\caption{Qubit-state discrimination accuracy results (\S~\ref{evaluation:discrimination}).}
\label{tab:model_accuracies}
\begin{tabular}{|c|c|c|c|c|c|c|}
\hline
\textbf{Design} & \textbf{Q1} & \textbf{Q2} & \textbf{Q3} & \textbf{Q4} & \textbf{Q5} & $\mathbf{F_{5Q}}$ \\ \hline
Baseline FNN & 0.968 & 0.746 & 0.941 & 0.945 & 0.970 & 0.909  \\ \hline
{QubiCML} & {0.968} & {0.731} & {0.895} & {0.936} & {0.957} & {0.892} \\ \hline
HERQULES\footnote{HERQULES exhibits significant variance between independent re-trainings. $F_{5Q}=0.905\pm 0.007$} & 0.969 & 0.740 & 0.925 & 0.942 & 0.970 & 0.905 \\ \hline
\projecttitle{}-T & 0.970 & 0.751 & 0.941 & 0.945 & 0.971 & 0.911 \\ \hline
\projecttitle{}-CNN & 0.971 & 0.747 & 0.941 & 0.947 & 0.971 & 0.911 \\ \hline
\end{tabular}
\end{table}

\begin{table}[t]
\footnotesize
\caption{Discrimination accuracy vs. trace length (\S~\ref{evaluation:discrimination}). }
\label{tab:model_accuracies_short}

\setlength{\tabcolsep}{4pt}
\begin{tabular}{|c|c|c|c|c|c|c|c|}
\hline
\textbf{Design}  &\textbf{Length} & \textbf{Q1} & \textbf{Q2} & \textbf{Q3} & \textbf{Q4} & \textbf{Q5} & $\mathbf{F_{5Q}}$\\ \hline
\multirow{3}{*}{HERQULES} 
& 750 & 0.831 & 0.721 & 0.922 & 0.895 & 0.969 & 0.863 \\ \cline{2-8}
& 500 & 0.513 & 0.663 & 0.915 & 0.631 & 0.936 & 0.713 \\  \cline{2-8}
& 250 & \cellcolor{red!25} 0.500 & \cellcolor{red!25} 0.550 & \cellcolor{red!25} 0.506 & \cellcolor{red!25} 0.500 & \cellcolor{red!25} 0.500 & \cellcolor{red!25} 0.511 \\  \hline
\hline

\multirow{3}{*}{{QubiCML}} & {750} & {0.960} & {0.725} & {0.905} & {0.936} & {0.963} & {0.892} \\ \cline{2-8}
& {500} & {0.928} & {0.707} & {0.915} & {0.918} & {0.962} & {0.880} \\  \cline{2-8} 
 & {250} &  {0.803} &  {0.655} &  {0.878} & {0.806} &  {0.806} & {0.786}  \\ \hline  \hline

\multirow{3}{*}{\projecttitle{}-CNN} 
& 750 & 0.964 & 0.736 & 0.940 & 0.943 & 0.970 & 0.906 \\ \cline{2-8}
& 500 & 0.939 & 0.715 & 0.934 & 0.927 & 0.968 & 0.892 \\ \cline{2-8}
& 250 & \cellcolor{green!25} 0.835 & \cellcolor{green!25} 0.664 & \cellcolor{green!25} 0.895 & \cellcolor{green!25} 0.833 & \cellcolor{green!25} 0.856 & \cellcolor{green!25} 0.812 \\ \cline{2-8}

\hline

\end{tabular}
\end{table}

\subsection{Classical Feedback Performance}
\label{evaluation:classical_feedback}

We evaluate \projecttitle{}'s classical feedback performance by (1) comparing the feedback latency of \projecttitle{} to Qubic, and (2) the impact of this latency on quantum applications.

\myparagraph{Branch instruction latency}
Table \ref{tab:latency_full_comparison} shows the branch instruction and total classical feedback latencies (excluding readout) for Qubic compared to \projecttitle{}. On the left, we increase the input size for the XOR or majority voting operation. We additionally include 128-bit operations that our controller supports at the same constant latency. Notably, Qubic incurs up to $32\times$ higher branch latency, which leads to $241\%$ higher feedback latency, for 32 qubits. Note that 701ns of classical feedback is comparable to measurement duration (Table \ref{tab:error_rates}).

\myparagraph{Feedback latency impact on fidelity}
Fig. \ref{fig:feedback_impact} shows the impact of feedback latency on the fidelities of 16-qubit constant-depth GHZ and long-range CNOT benchmarks. Here, we only model the thermal relaxation errors caused by qubit idling, for the sake of fairness. On the x-axis, we increase the number of instances of the GHZ state or CNOT within a higher-level application (e.g., MECH). Notably, \projecttitle{} achieves, on average, $57.3\%$ and $37.8\%$ higher fidelity for the two applications. Last, \projecttitle{} improves MECH's effective circuit depth by  $\sim 7\times$ compared to Qubic (Fig. \ref{fig:mech_analysis} (c)).

\takeaway{
\projecttitle{}'s constant feedback latency is up to $70\%$ lower than Qubic's, which leads to $37.8\%$-$57\%$ higher fidelity and up to $7\times$ lower depth for quantum applications, on average.
}

\subsection{Qubit-State Discrimination Performance}
\label{evaluation:discrimination}

We evaluate the performance of our qubit-state discriminator models w.r.t. (1) discrimination accuracy, (2) inference latency and FPGA resource utilization, and (3) impact on application fidelity.

\myparagraph{Discrimination accuracy with long traces}
Table \ref{tab:model_accuracies} shows the fidelities of the baseline FNN, QubiCML, HERQULES, and the \projecttitle{} discriminators, for 1$\mu s$ readout traces. We show the individual qubit fidelities, and the geometric mean fidelity for all five qubits.

\begin{figure*}[t]
    \centering
    \setlength{\fboxrule}{0.5pt}
    \includegraphics[width=\textwidth]{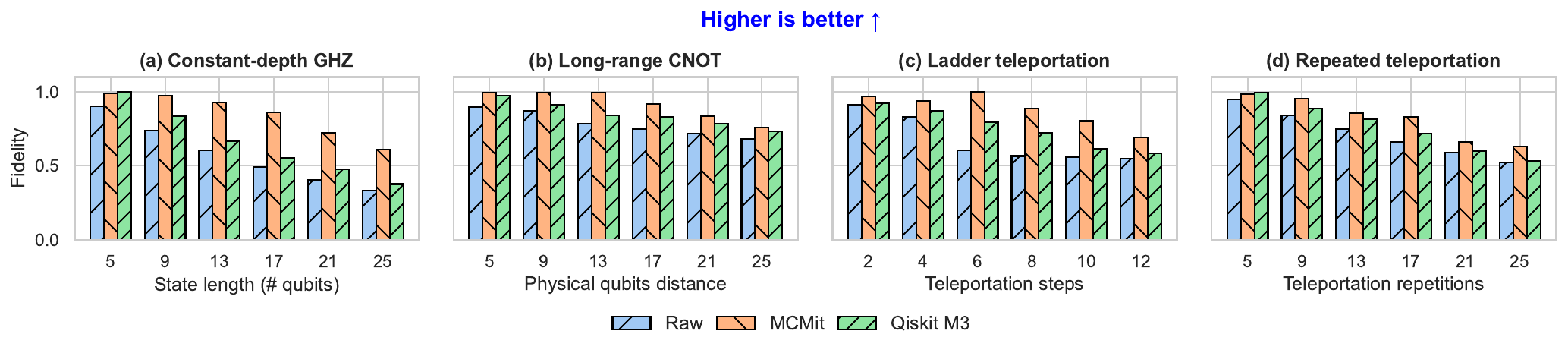}
    \caption{
    \projecttitle{} software error mitigation impact on fidelity (\S~\ref{evaluation:software}). \textit{
     \projecttitle{} achieves 30.47\% and 17.78\% higher fidelity than raw execution and Qiskit M3, respectively, on average. 
    }  
    }
    \label{fig:software_eval}
\end{figure*}

Notably, both \projecttitle{}-T and \projecttitle{}-CNN are 2.1\% more accurate than the baseline FNN, 17.5\% more accurate than QubiCML, and 6.3\% more accurate than HERQULES. Moreover, our models require no hand-crafted feature extraction or pre-classifiers, and achieve substantially higher accuracy for short readout traces, as we show next.

\begin{table}[t]
  \centering
  \footnotesize
  \caption{Mean absolute cross-fidelity $\langle|F^{\mathrm{CF}}_{ij}|\rangle$ by
  qubit separation $|i-j|$ (lower is better; perfect crosstalk immunity $=0$).
  All-qubit joint readout, full $1000$\,ns window.}
  \label{tab:crossfid}
  \begin{tabular}{lcccc}
    \toprule
    Design & $|i-j|{=}1$ & $|i-j|{=}2$ & $|i-j|{=}3$ & $|i-j|{=}4$ \\
    \midrule
    HERQULES  & \textbf{0.0033} & 0.0111 & \textbf{0.0015} & \textbf{0.0005} \\
    MCMit-CNN & 0.0035 & \textbf{0.0077} & 0.0020 & 0.0011 \\
    \bottomrule
  \end{tabular}
\end{table}

\begin{table}[t]
  \centering
  \footnotesize
  \caption{Discrimination accuracy as the number of simultaneously measured qubits $N$ increases,
  averaged over all $\binom{5}{N}$ active-qubit subsets.}
  \label{tab:scaling}
  \begin{tabular}{lccccc}
    \toprule
    Design & $N{=}1$ & $N{=}2$ & $N{=}3$ & $N{=}4$ & $N{=}5$ \\
    \midrule
    HERQULES  & 0.922 & 0.916 & 0.911 & 0.908 & 0.905 \\
    \projecttitle{}-CNN & \textbf{0.924} & \textbf{0.919} & \textbf{0.916} & \textbf{0.913} & \textbf{0.910} \\
    \bottomrule
  \end{tabular}
\end{table}

\myparagraph{Discrimination accuracy with short traces}
Table \ref{tab:model_accuracies_short} shows the accuracies for HERQULES, QubiCML, and \projecttitle{}-CNN for traces spanning 250–750ns. Interestingly, \projecttitle{}-CNN is substantially more robust to short readout traces compared to the baselines. Starting at 750ns (25\% faster readout), \projecttitle{}-CNN is 31.4\% more accurate than HERQULES; the latter drops fidelity by 4.6\% while \projecttitle{}-CNN only by 0.5\% (8.4$\times$ less). At 500 ns length (50\% faster readout), \projecttitle{}-CNN achieves 62.4\% higher accuracy than HERQULES (78.8\% excluding Qubit 2); HERQULES' fidelity drops by 21.2\% while \projecttitle{}-CNN by 2.1\% (10.1$\times$ less). At 250ns length (75\% faster readout), \projecttitle{}-CNN is 61.6\% more accurate (70.8\% excluding Qubit 2); HERQULES' fidelity drops by 43.5\% while \projecttitle{}-CNN only by 10.9\% (4.0$\times$ less). Last, \projecttitle{}-CNN achieves 12.9\% (750ns), 9.6\% (500ns), and 12.3\% (250ns) lower infidelity than QubiCML, on average.

\myparagraph{Crosstalk mitigation}
We quantify crosstalk-induced cross-fidelity as a function of qubit separation/distance, $k$, i.e., the spurious change in a qubit's readout fidelity caused by the state of another qubit $k$ sites away. HERQULES \cite{maurya2023scaling} reports this as a fidelity distance (deviation from the crosstalk-free baseline), so larger values mean stronger inter-qubit leakage and smaller values mean the classifier is more robust to neighboring-qubit activity. Table \ref{tab:crossfid} shows that the two are essentially on par across all separations: they are nearly identical at short range ($k{=}1$: 0.0035 vs 0.0033), our CNN is more robust at $k{=}2$ (0.0077 vs 0.0111, about $1.4\times$ smaller), while HERQULES is marginally better at $k{=}3$ (0.0020 vs 0.0015) and $k{=}4$ (0.0011 vs 0.0005). Both classifiers share the same non-monotonic dependence on $k$, peaking mildly at $k{=}2$, and reach comparably low cross-fidelity overall.

\myparagraph{Accuracy scaling with simultaneously measured qubits}
We evaluate how discrimination fidelity scales with the number of simultaneously measured qubits, $N$, by averaging the single-qubit assignment fidelity over all $\binom{5}{N}$ subsets of driven qubits while holding the remaining qubits in the ground state. Table \ref{tab:scaling} shows that \projecttitle{}-CNN exhibits both a shallower slope and a higher accuracy at every load as the multiplexed register grows from $N{=}1$ to $N{=}5$: \projecttitle{}-CNN falls by $1.4$ percentage points ($0.924\rightarrow0.910$, a slope of $0.33$ pp per added qubit) while HERQULES falls by $1.7$ percentage points ($0.922\rightarrow0.905$, $0.44$ pp per qubit). \projecttitle{}-CNN stays ahead at every $N$, reaching $0.910$ versus HERQULES' $0.905$ at full $N{=}5$ while matching HERQULES in crosstalk robustness as the register fills.

\begin{figure}[t]
    \centering
    \includegraphics[width=\columnwidth]{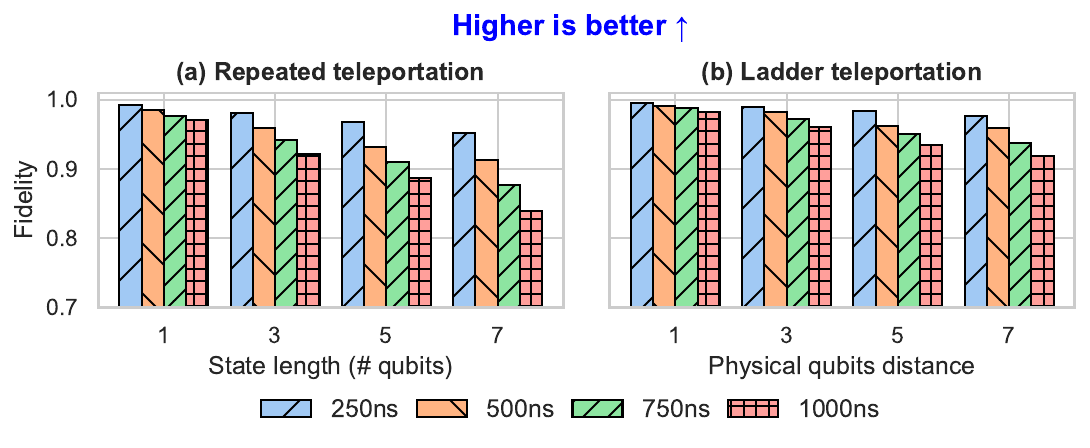}
    \caption{Readout duration impact on fidelity (\S~\ref{evaluation:discrimination}).
    \textit{
    The x-axis shows the number of teleportation steps. A 250ns readout achieves 6\% higher fidelity than that of 750ns, on average.
    }
    }
    \label{fig:mcm_impact}
\end{figure}

\begin{table}[t]
\centering
\footnotesize

\caption{Discriminator FPGA resource utilization and latency on XCU280 (8-bit quantization, 500 MHz clock).}
\label{tab:fpga_utilization}

\setlength{\tabcolsep}{4pt}
\begin{tabular}{|c|c|c|c|c|c|}
\hline
\textbf{Design} & \textbf{DSP (\%)} & \textbf{BRAM (\%)} & \textbf{FF (\%)} & \textbf{LUT (\%)} & \textbf{Latency (ns)} \\ \hline

\projecttitle{}-T  & 23.6 & 9.0 & 1.9  & 29.1 & 80 \\  \hline
\projecttitle{}-CNN
  & 0 & 0 & 0.10 & 0.33 & 70 \\ 
\hline
{HERQULES} & {0.75} & {0.75} & {0.41}  & {4.2} & {40} \\
\hline
{QubiCML}  & {15} & {0.94} & {2.82}  & {1.87} & {54} \\
\hline

\end{tabular}
\end{table}

\myparagraph{FPGA resource utilization and inference latency}
Table \ref{tab:fpga_utilization} shows the XCU280 FPGA resource utilization and the inference latency for \projecttitle{} and the baselines, for five-qubit state discrimination and 1$\mu$s readout duration. Although the \projecttitle{} transformer incurs higher overheads compared to the baselines, the \projecttitle{}-CNN achieves 92\% and 82.3\% lower LUT utilization, and 75\% and 96.4\% lower FF utilization compared to HERQULES and QubiCML, respectively.

\myparagraph{MCM latency impact on fidelity}
Fig. \ref{fig:mcm_impact} shows the impact of readout duration on the fidelity of quantum teleportation circuits. Both circuits scale in depth with the increasing number of teleportation steps (x-axis). Here, we only model the thermal relaxation errors caused by qubit idling, for the sake of fairness. Notably, a readout length of 250ns (500ns) achieves $5.8\%$ ($2.8\%$) higher fidelity than a readout length of 750ns, on average. Last, \projecttitle{} (250ns) achieves $30.9\%$ lower depth than HERQULES (750ns) on MECH (Fig. \ref{fig:mech_analysis} (c)).

\takeaway{
\projecttitle{} achieves $62\%$ and $12\%$ higher accuracy for 250ns readout compared to HERQULES and QubiCML, respectively, leading to $30.9\%$ lower depth for DQC applications, on average.
}

\subsection{Software Mitigation Performance}
\label{evaluation:software}
We evaluate the effectiveness of the \projecttitle{} software error mitigation by (1) running the four micro-benchmarks on \texttt{ibm\_fez} and measuring fidelity, (2) showing the impact of MCM error improvement on MECH and the surface QEC code.

\myparagraph{Fidelity improvement}
Fig. \ref{fig:software_eval} shows the fidelity evaluation for our four micro-benchmarks. For the constant-depth GHZ state and long-range CNOT benchmarks, we increase the state (distance) length in number of qubits. In the case of quantum teleportation, we increase the number of consecutive teleportations (in a ladder or sequential manner, respectively). ``Raw'' indicates completely unmitigated results, and the \projecttitle{} and Qiskit M3 results do not use other error mitigation techniques, for fairness. Each experiment runs for 10,000 shots on the same calibration cycle of \texttt{ibm\_fez}.

Notably, \projecttitle{} achieves 30.47\% and 17.78\% higher fidelity than raw execution and Qiskit M3, respectively, on average. Specifically, the improvements for the four applications are: 56.9\% and 27.84\%, 16.5\% and 8\%, 33.5\% and 17.8\%, and 14.6\% and 8\%. We observe the largest fidelity improvement in constant-depth GHZ, as it requires the most MCMs and classical feedback operations, which are effectively eliminated by our dynamic circuit simplification stage. Also, the GHZ state is protected by the parity checks added by the measurement hardening stage. In contrast, the repeated teleportation circuit achieves the smallest improvement over Qiskit M3, since it scales only with depth and circuit width; thus, our measurement hardening stage is less effective than wide circuits.

\begin{figure*}[t]
    \centering
    \setlength{\fboxrule}{0.5pt}
    \includegraphics[width=\textwidth]{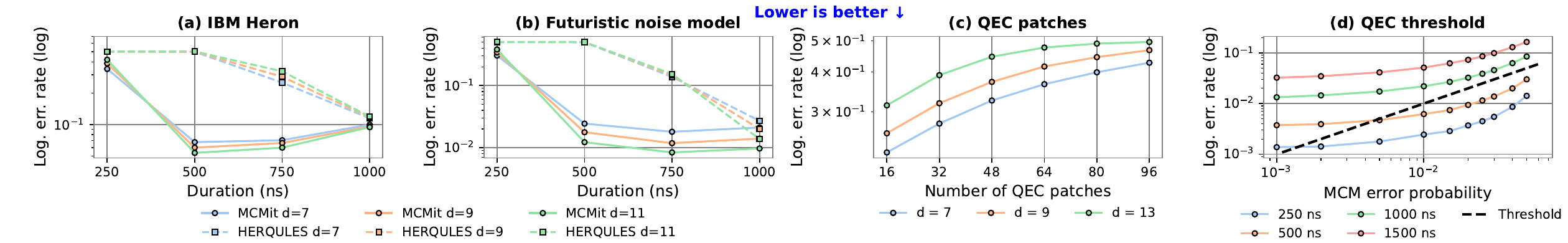}
    \caption{
    Impact of varying readout duration and fidelity on logical error rate. \emph{(a) and (b) compare MCMit and HERQULES under two noise models, (c) shows the additional latency in MCMit readout as the number of QEC patches d=7 grows, (d) explores different MCM durations and error rates, highlighting regimes where errors are effectively suppressed on surface code patches d=7.}}
    
    \label{fig:qec_eval}
\end{figure*}

Last, Table \ref{tab:elimination_runtime} shows the compilation runtime to statically eliminate MCMs. The runtime is dominated not by the number of MCMs but by the cost of classically simulating the circuit's intermediate state: it stays negligible for the teleportation variants (small, sparse states) and still scales for GHZ, but grows sharply for the long-range CNOT, whose protocol transiently builds a dense $\sim 2^{n/2}$-amplitude state. Thus the method's practical limit is set by the circuit's classical simulability rather than by its MCM count.

\takeaway{
The \projecttitle{} software error mitigation achieves $\sim30\%$ and $\sim18\%$ higher fidelity than unmitigated and Qiskit M3 results, respectively, on average, with up to $\sim27.84\%$ improvement over Qiskit M3 for the constant-depth GHZ circuit.
}

\subsection{QEC Performance}
\label{eval:qec}

\begin{table}[t]
\centering
\footnotesize
\caption{Compilation runtime to eliminate MCMs.}
\label{tab:elimination_runtime}
\begin{tabular}{lcccc}
\hline
\textbf{Application} & \textbf{small $n$} & \textbf{runtime} & \textbf{large $n$} & \textbf{runtime} \\
\hline
Constant-depth GHZ & 5 & $0.003$\,s & 25 & $6.8$\,s \\
Long-range CNOT    & 5 & $0.002$\,s & 25 & $2272$\,s ($\sim$38\,min) \\
Repeated teleportation   & 5 & $0.017$\,s & 25 & $0.289$\,s \\
Ladder teleportation    & 5 & $0.013$\,s & 12 & $0.113$\,s \\
\hline
\end{tabular}
\end{table}

We evaluate the impact of MCMit-CNN’s readout fidelity on logical error rates by comparing it to HERQULES \cite{maurya2023scaling} under two noise models: (1) a realistic IBM Heron device \cite{IBM_roadmap_2025}, and (2) a futuristic model where gate errors are downscaled $10\times$ and decoherence times prolonged $3\times$.
The setup is explained in \S~\ref{motivation:qec}.
We also examine how classical feedback, resulting from additional QEC patches, affects logical error rates and identify regimes of readout latency and fidelity that support effective fault-tolerant execution.

\myparagraph{Logical error rate improvement} Fig. \ref{fig:qec_eval} (a) and (b) show the logical error rate (LER) achieved using the fidelity and readout duration of MCMit and HERQULES, averaged over $d=\{7,9,11\}$. \projecttitle{} achieves lower LERs than HERQULES at every readout length and in both noise regimes. In the current noise regime, \projecttitle{} reduces LERs over HERQULES by 1.2--1.5$\times$ (250ns), 7.4--9.4$\times$ (500ns), and 1.2--1.3$\times$ (1$\mu$s). Within \projecttitle{}, a fast 500ns readout is the sweet spot, lowering LERs by $32$--$43\%$ ($d{=}7/9/11$) compared to 1$\mu$s readout.
In the futuristic regime, the gap widens sharply: \projecttitle{} reduces LERs over HERQULES by 1.3--1.6$\times$ (250ns) and 20--41$\times$ (500ns), where HERQULES collapses to a near coin-flip LER ($\sim0.50$) at 500ns while \projecttitle{} stays at $0.012$--$0.024$. In this low-error regime, the optimal readout instead shifts to 750ns (a 7.6--18$\times$ reduction over HERQULES), and a 500ns readout is $15$--$27\%$ higher than 1$\mu$s.

{\myparagraph{Impact of classical delay from additional patches}
For lattices exceeding 16 QEC patches, each additional patch introduces 16 ns of classical feedback overhead. Fig. \ref{fig:qec_eval} (c) illustrates the impact of this delay on logical error rate using MCMit readout parameters under the IBM Heron noise model. The degradation trend is consistent across all code distances, notably, with larger distances reaching failure earlier, indicating increased sensitivity to accumulated classical delay.}

{\myparagraph{Exploration of varying latency and error regimes} Fig. \ref{fig:qec_eval} (d) presents logical error rates under different combinations of readout latency and error. The gray line marks the threshold where the logical error rate matches or falls below the MCM error probability, indicating regimes that effectively suppress errors. Our results highlight the critical importance of readout improvements, with 500 ns showing the most favorable trade-off between latency and error suppression.}

\takeaway{{MCMit outperforms HERQULES at every readout length, achieving up to $9.4\times$ and $41\times$ lower logical error rates at 500ns readout in the current and futuristic noise regimes, respectively.}}

\section{Related Work}
\label{related_work}

\myparagraph{Readout error mitigation}
Readout error mitigation through statistical and machine learning techniques is an active area of research  \cite{das2021jigsaw, tannu2019mitigating, tannu2022hammer, berg2022model, bravyi2021mitigating, kim2022quantum, alistair2021qubit, yang2022efficient, aasen2024readout, maciejewski2020mitigation}. However, these methods operate at the level of final execution results, adjusting measurement outcomes post-execution to approximate the correct output probabilities, and therefore  are only  effective for terminal measurements, not MCMs.
In contrast, our work holistically mitigates MCM-induced errors by focusing on MCM and branching errors, and dynamic circuit latency, which causes decoherence errors.

\myparagraph{FPGA-based controller frameworks}
The development of high-performance, real-time FPGA-based quantum controllers is an active and constantly evolving area of research \cite{xu2021qubic, wuwei2025artery, diguglielmo2025end, messaoudi2020a, yang2022FPGA}.
However, to our knowledge, existing open-source solutions offer limited support for fast, scalable classical feedback logic, such as multi-qubit conditional branching, and hardware acceleration, like LUTs for constant-time operations critical to QEC and DQC. In contrast, \projecttitle{} implements both a constant-latency branch operation and LUTs for complex ALU operations.

\myparagraph{Qubit-state discrimination}
High-accuracy qubit-state discrimination remains an open problem, with various approaches proposed to maximize readout fidelity \cite{vora2024ml, guo2025klinq, mude2025efficient, maurya2023scaling, lienhard2022deep, giortamis2026oraqleempiricalanalysisqubit}.
While some of these models are effective for \textit{long} readout durations, they underperform for short.
To this end, we introduce two discriminator designs: a transformer that leverages self-attention to capture non-local temporal dynamics even under short readout durations, and a lightweight residual CNN that exploits local temporal correlations directly from raw downsampled traces without hand-crafted feature extraction or pre-classifiers.

\myparagraph{DQC systems software}
DQC compilers and runtimes focus on minimizing inter-QPU communication \cite{wu2022autocomm, wuqucomm2023, escofet2025revisiting, zhang2024optimizing, cambiucci2023hypergraphic, martinez2019automated, baker2020time, dadkhan2022reordering, davarzani2020dynamic, diadamo2021distributed, ferrari2021compiler, haner2021distributed, tornow2025qvm, giortamis2025qos, giortamis2025qonductor, tornow2024qtpu}. While these approaches reduce the frequency of cross-QPU communication, the number of MCMs remains proportionally larger than the number of remote gates and continues to impose a significant fidelity and latency cost (\S~\ref{motivation:dqc}). In contrast, \projecttitle{} focuses on removing MCMs and their feedback operations entirely, when possible, and mitigating their errors otherwise.

\myparagraph{Compilers for QEC} Existing QEC compilers decompose quantum states and gates into fault-tolerant primitives, and manage their mapping and scheduling, primarily for surface code mechanisms such as lattice surgery~\cite{Paler_2017, Herr_2017, Watkins_2024, Leblond_2023, Liu_2023, Beverland_2022} or braiding~\cite{10.1109/MICRO.2018.00072, swierkowska2025eccentrice}. Consequently, they operate at a higher level of abstraction, rather than directly addressing physical-level error mechanisms.

\section{Conclusion}
\label{conclusion}

In this work, we presented \projecttitle{}, a compiler–controller co-design that addresses the critical bottlenecks of MCM error and feedback latency in DQC and QEC. Our software mitigates errors through static analysis and stochastic branching decisions, while our controller hardware introduces a constant-latency multi-control branch for fast feedback and two high-fidelity discriminators for fast, accurate qubit-state readout.
As MCM-heavy dynamic circuits become central to both QEC and DQC, we believe this kind of tight hardware-software co-design will be essential to keeping classical control from becoming the dominant source of error and latency.

\section*{Acknowledgments}
This work is funded by the Bavarian State Ministry of Science and the Arts as part of the Munich Quantum Valley (MQV), grant number 6090181. B.L. acknowledges support from the German Federal Ministry of Research, Technology and Space (BMFTR) through the program EQuIPS (Grant No. 13N17232).

\bibliographystyle{IEEEtran}
\bibliography{bibliography}

\end{document}